\documentclass{aa}
\usepackage[varg]{txfonts}
\usepackage{graphicx}
\usepackage{natbib}
\bibpunct{(}{)}{;}{a}{}{,} 
\usepackage{hyperref}
\usepackage{amsmath}
\usepackage{amssymb}
\hypersetup{colorlinks=true,linkcolor=blue,citecolor=blue,urlcolor=blue} 

\newcommand{\methane}{$\mathrm{CH_4}$}
\newcommand{\methyla}{$\mathrm{C_3H_4}$}
\newcommand{\cyano}{$\mathrm{C_2N_2}$}
\newcommand{\diacety}{$\mathrm{C_4H_2}$}

\begin{document}

\title{Seasonal evolution of \cyano, \methyla, and \diacety~abundances in Titan's lower stratosphere}
\titlerunning{Seasonal evolution of \cyano, \methyla, and \diacety}

\author{M. Sylvestre \inst{\ref{inst1}} 
\and N.A. Teanby \inst{\ref{inst1}}
\and S. Vinatier \inst{\ref{inst2}}
\and S. Lebonnois \inst{\ref{inst4}} 
\and P. G. J. Irwin\inst{\ref{inst3}}
}

\institute{School of Earth Sciences, University of Bristol, Wills Memorial Building, Queen's Road,\\
Bristol BS8 1 RJ, UK \label{inst1}
\and LESIA, Observatoire de Paris, PSL Research University, CNRS, Sorbonne Universités, UPMC Univ. Paris 06, Univ. Paris Diderot, Sorbonne Paris Cité, 5 Place Jules Janssen, 92190 Meudon, France \label{inst2}
\and LMD, CNRS, IPSL, UMR 8539, 4 Place Jussieu, F-750005, Paris France \label{inst4}
\and Atmospheric, Oceanic, \& Planetary Physics, Department of Physics, University of Oxford, Clarendon Laboratory, Parks Road, Oxford OX1 3PU, UK \label{inst3}
}

\abstract{}{We study the seasonal evolution of Titan's lower stratosphere (around 15~mbar) in order to better understand the atmospheric dynamics and chemistry in this part of the atmosphere.}{We analysed Cassini/CIRS far-IR observations from 2006 to 2016 in order to measure the seasonal variations of three photochemical by-products: \diacety, \methyla,~and \cyano.}{We show that the abundances of these three gases have evolved significantly at northern and southern high latitudes since 2006. We measure a sudden and steep increase of the volume mixing ratios of \diacety, \methyla, and \cyano~at the south pole from 2012 to 2013, whereas the abundances of these gases remained approximately constant at the north pole over the same period.  At northern mid-latitudes, \cyano~and \diacety~abundances decrease after 2012 while \methyla~abundances stay constant. The comparison of these volume mixing ratio variations with the predictions of photochemical and dynamical models provides constraints on the seasonal evolution of atmospheric circulation and chemical processes at play.}{}{}

\maketitle

\section{Introduction} 

        Titan's atmosphere undergoes a rich photochemistry, initiated by the dissociation of its most abundant constituents, $\mathrm{N_2}$ and \methane. In the thermosphere and the ionosphere (at altitudes above 600~km or the 0.0001~mbar pressure level), these molecules are dissociated by solar UV and EUV photons, energetic photoelectrons, and high energy electrons from Saturn's magnetosphere \citep{Wilson2004,vuitton2012}. Radicals and ions produced by these photodissociations react together and form hydrocarbons, nitriles, and eventually organic hazes. These species are then destroyed by photolysis or further chemical reactions in the upper and middle atmosphere, or they condense in the lower part of the stratosphere (at altitudes inferior to 100~km or at pressures superior to 10~mbar). As Titan's obliquity is $26.7^{\circ}$, its atmosphere undergoes significant seasonal variations of insolation which are expected to affect the abundances of the photochemical by-products. In addition, this seasonal forcing affects atmospheric dynamics, which transports minor atmospheric constituents \citep{Teanby2008,Teanby2012,Vinatier2015}. Hence, measuring the meridional and vertical distributions of the various photochemical species is a way to better understand the photochemical and dynamical processes in Titan's atmosphere.\\

        Abundances of hydrocarbons and nitriles and their temporal evolution have been measured in various studies, especially since the beginning of the Cassini mission in 2004. This spacecraft has provided thirteen years of observations of Titan at different wavelength ranges, enabling us to monitor the seasonal evolution of its atmosphere throughout its northern winter and spring. During Titan's northern winter, limb and nadir observations performed with the infrared spectrometers Voyager 1/IRIS in November 1980 (Infrared Instrument, \citet{Hanel1981}) and Cassini/CIRS (Composite InfraRed Spectrometer, \citet{Flasar2004}) between 2004 and 2008 showed that many species such as acetylene ($\mathrm{C_2H_2}$), diacetylene (\diacety), or cyanoacetylene ($\mathrm{HC_3N}$) exhibited an enrichment at high northern latitudes \citep{Kunde1981,Coustenis1991,Coustenis2007,Coustenis2010,Teanby2008b,Vinatier2010,Bampasidis2012}. This was attributed to the atmospheric circulation which took the form of a single pole-to-pole Hadley cell, and more specifically to the subsiding branch of this cell, located above the winter pole and bringing photochemical species from their production level to the stratosphere \citep{Teanby2009b,Vinatier2010}. After the equinox (August 2009), CIRS measurements analysed by \citet{Teanby2012, Coustenis2013, Vinatier2015, Coustenis2016} revealed that the vertical and meridional distributions of these gases have changed significantly, especially above the south pole where the abundances of photochemical by-products strongly increased after 2011. This was interpreted as a subsidence above high southern latitudes, due to the reversal of the pole-to-pole Hadley cell.  At northern latitudes, \citet{Coustenis2013} found a decrease in trace gas abundances between 2010 and 2012. \\

        While mid-infrared CIRS observations provide information between 10~mbar and 0.001~mbar for limb observations, and from 10~mbar to 0.5~mbar for nadir observations, far-infrared CIRS spectra mainly probe the lower part of the stratosphere, at pressure levels between 10~mbar and 20~mbar. For instance, this type of observation allowed \citet{Teanby2009} to measure the meridional distribution of diacetylene (\diacety) and methylacetylene (\methyla) in the lower stratosphere during winter. Furthermore, unlike the CIRS mid-infrared observations, far-infrared spectra can be used to measure the distribution of cyanogen (\cyano), thus providing additional information about the chemistry of Titan's stratosphere. For instance, \citet{Teanby2009} compared the enrichment in \cyano~and other nitriles and hydrocarbons at the north pole during winter, and suggested that nitriles undergo an additional loss process compared to the hydrocarbons with similar lifetimes, as proposed by \citet{Yung1987}.\\

        In this paper, we use nadir far-infrared spectra from Cassini/CIRS to measure the meridional distributions of diacetylene (\diacety), methylacetylene (\methyla), and cyanogen (\cyano) from 2006 to 2015. The data we present cover the whole latitude range and were acquired throughout the Cassini mission. It allows us to monitor precisely the seasonal evolution of the distributions of \cyano, \methyla, and \diacety~in the lower stratosphere, and thus to complete the previous studies by giving insights on the atmospheric dynamics and photochemistry of Titan's lower stratosphere. 

\section{Observations}
\label{sect_obs}

                        \begin{figure}[!h]
                                \centering
                                \includegraphics[width=1\columnwidth]{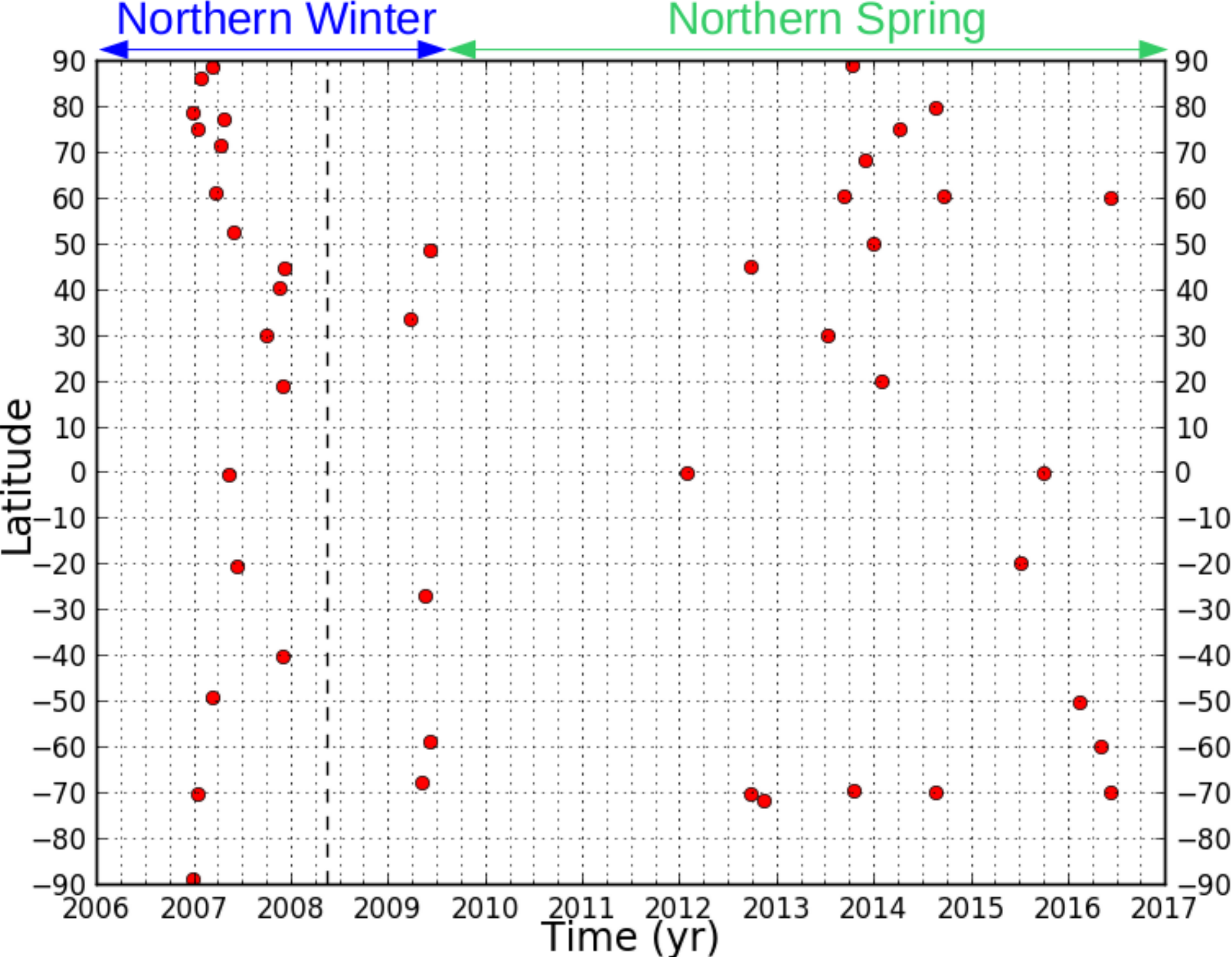}
                                \caption{Spatial and temporal distribution of the FP1 data analysed in this paper. They cover all the latitudes, with a spatial field of view of $20^{\circ}$. Observations are available for different times throughout northern winter and spring. Their temporal distribution is controlled by the orbits of the Cassini spacecraft.} 
                                \label{fig_repartobs}
                        \end{figure}
                        
                        Cassini Composite InfraRed Spectrometer (CIRS, \cite{Flasar2004}) is a  Fourier transform spectrometer composed of three focal planes which operate in different wavenumber ranges. The focal plane FP1 probes the spectral range 10 - 600$~\mathrm{cm^{-1}}$ (17 - 1000$~\mathrm{\mu m}$) and is made of a single circular detector with an angular resolution of 3.9 mrad. The focal planes FP3 and FP4 respectively measure spectra in  600 - 1100$~\mathrm{cm^{-1}}$ (9 - 17 $~\mathrm{\mu m}$) and  1100 - 1400$~\mathrm{cm^{-1}}$ (7 - 9$~\mathrm{\mu m}$). Both are composed of a linear array of ten detectors with an angular resolution of 0.273~mrad per detector.\\ 
                
                        In this study, we analyse nadir spectra acquired from FP1, with an apodised  spectral resolution of 0.5$~\mathrm{cm^{-1}}$, in order to resolve the spectral signatures of \cyano, \methyla, and \diacety. We also exploit FP4 spectra acquired at the same resolution in order to measure temperature with the $\nu_4$ \methane~band ($1304~\mathrm{cm^{-1}}$). These observations were made in `sit-and-stare' geometry, where each detector of FP1 and FP4 probes the same latitude and longitude throughout the acquisition, with a total integration time comprised between 1h30 and 4h30. During these observations, the average spatial field of view is $20^{\circ}$ of latitude for the single detector of FP1, $2^{\circ}$ for each FP4 detector, and $15-20^{\circ}$ for the whole FP4 array, depending on its orientation.\\ 
        
                        The datasets used in this study are summarised in Table \ref{table_obs}. We selected data covering the whole latitude range, acquired from 2006 to 2016 (see fig. \ref{fig_repartobs}), in order to get an overview of the seasonal evolution from northern winter to mid-spring. For each dataset, 100 to 330 spectra were acquired with FP1, and 500 to 1650 with FP4. We use the photometric calibration provided by the CIRS team (version DS4000) which corrects the effects of sky background and thermal noise of the detectors more effectively than the standard calibration. Some of the datasets acquired during northern winter (before 2008, see table \ref{table_obs}) have already been presented in \cite{Teanby2009}, but as we use a different calibration version, we reanalyse them, to ensure a consistent comparison between these data and the other data presented in this study.\\
        
\begin{sidewaysfigure*}
                                        \label{fig_ex_spectres}
                                \begin{center}
                                        \includegraphics[width=0.37\columnwidth]{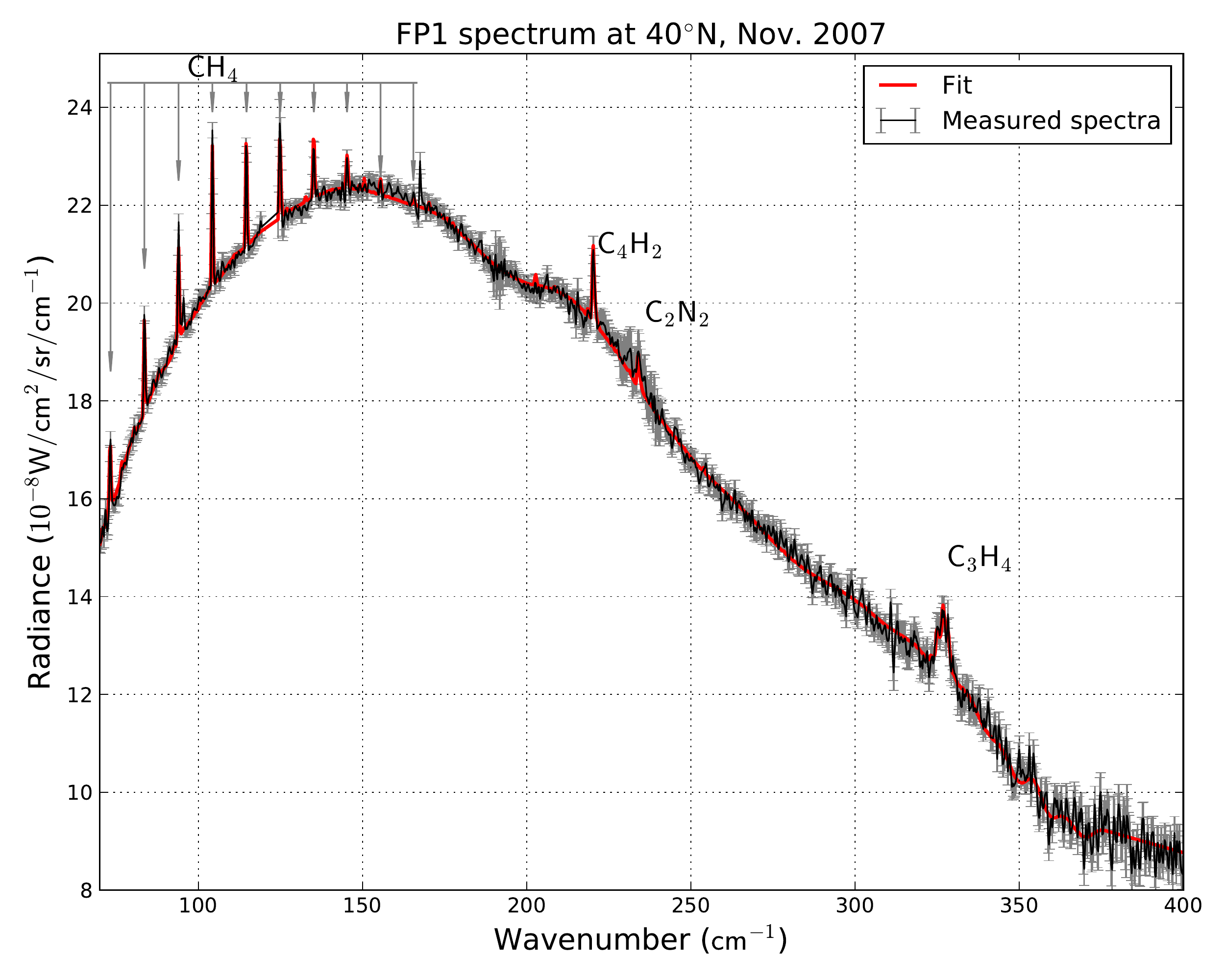}
                                        \includegraphics[width=0.37\columnwidth]{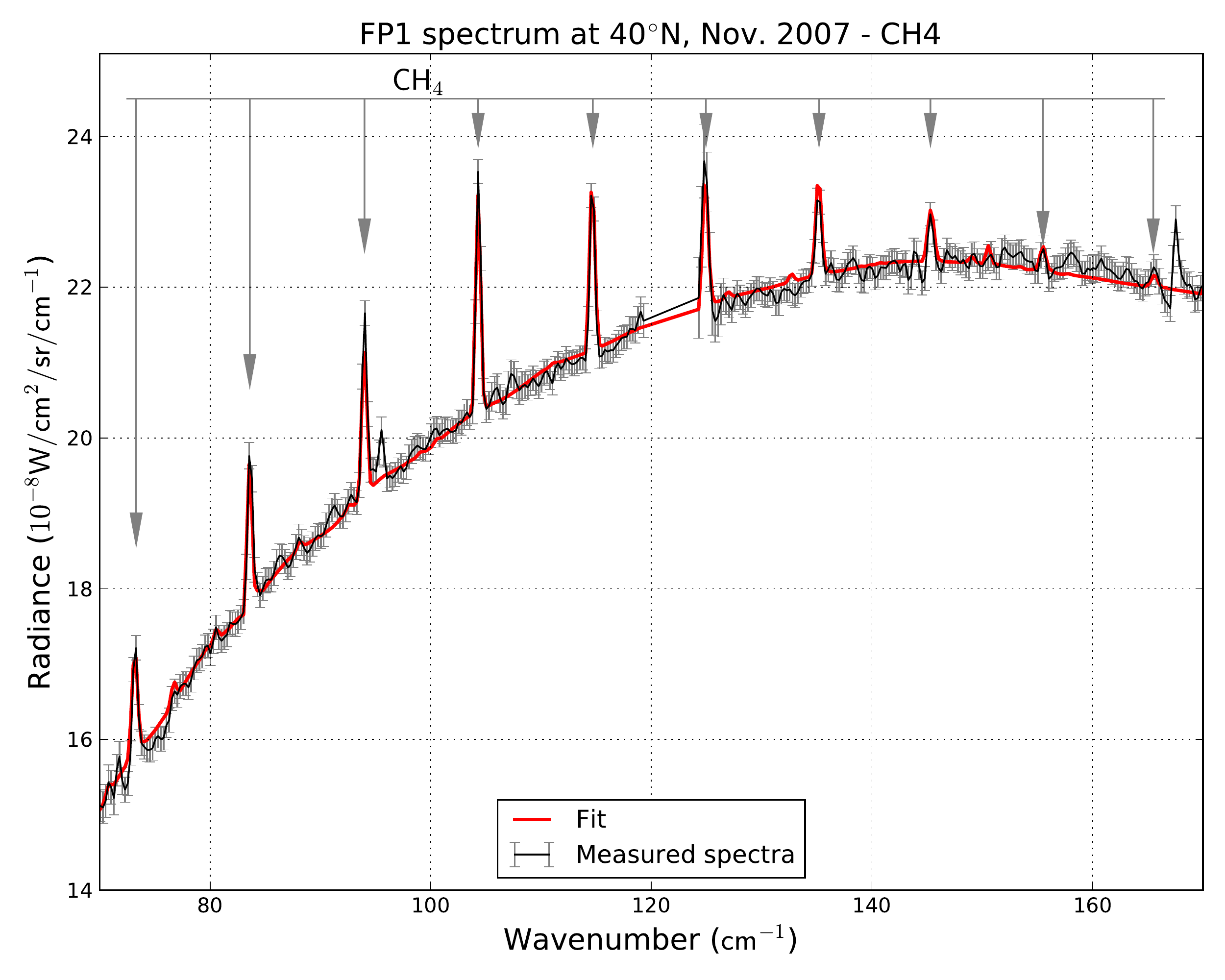}

                                        \includegraphics[width=0.37\columnwidth]{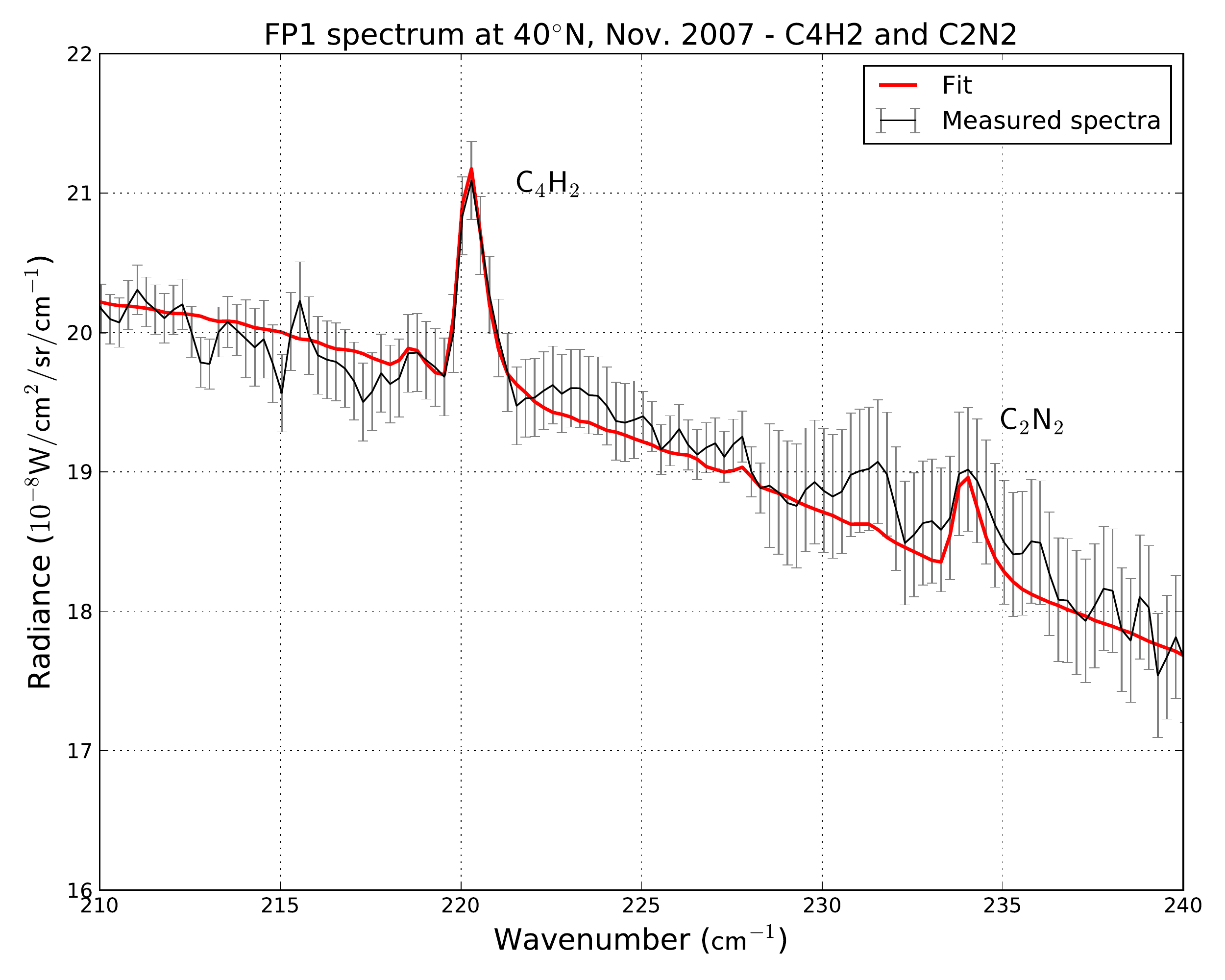}
                                        \includegraphics[width=0.37\columnwidth]{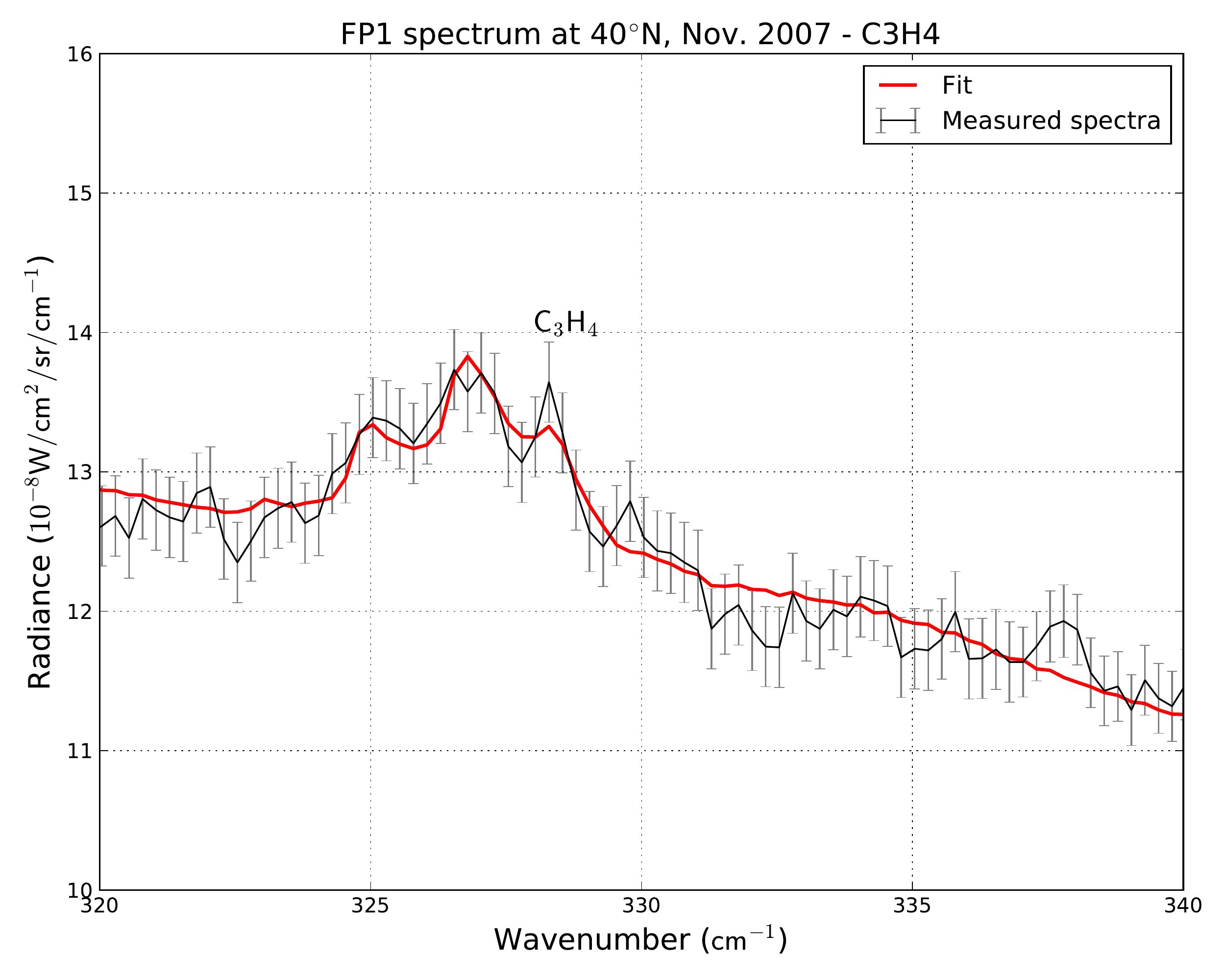}              
                                                \caption{Example of  FP1 spectrum (after average). The top left panel shows the whole spectrum while the three other panels are a close-up around the relevant spectral bands. Measured spectra are in black. Red solid lines show the synthetic spectra calculated during the retrieval process. Data were acquired at a spectral resolution of $0.5~\mathrm{cm^{-1}}$. Error bars of the FP1 spectrum have been corrected because the initial error bars were too small with respect to the radiance variations caused by noise (see sections \ref{sect_obs} and \ref{sect_corr_err}). \diacety, \cyano, and \methyla~bands are visible, and allow us to retrieve the volume mixing ratios of these species. The \methane~bands in FP1 are used to retrieve the temperature profile between 20~mbar and 10~mbar.} 
                                        \label{fig_ex_spectre}                  
                                \end{center}
                        \end{sidewaysfigure*}

                        For each FP1 and FP4 dataset, all the spectra acquired are averaged together in order to improve the signal-to-noise ratio by $\sqrt{N}$ (with $N$ the number of averaged spectra). The average signal-to-noise ratio reaches 115 at $220~\mathrm{cm^{-1}}$ (spectra acquired with FP1) and 160 at $1300~\mathrm{cm^{-1}}$ (FP4 spectra). Figure \ref{fig_ex_spectres} shows an example of spectrum obtained after averaging 271 FP1 spectra measured at $40^{\circ}$N in November 2007.\\  
                                                
\section{Analysis}
        \label{sect_analyse}
                        
        In order to retrieve the abundances of \diacety, \cyano, and \methyla,  we use the constrained non-linear inversion code NEMESIS \citep{Irwin2008}. Retrievals are performed following an iterative process based on the generation of synthetic spectra from a reference atmosphere, and the minimisation of a cost function in order to find the value of the retrieved parameter which provides the best fit of the measured spectrum. \\
                                                                
                \subsection{Reference atmosphere}
                         Our reference atmosphere extends from 0~km ($\sim1438$~mbar) to 780~km ($\sim~1\times10^{-5}$~mbar). The gases included in this study, their abundances, and the studies in which they were measured are detailed in table \ref{tab_atm}.  These measurements were performed using data from Cassini/CIRS \citep{Nixon2012,Cottini2012,Teanby2009,Coustenis2016}, Huygens/GCMS \citep{Niemann2010}, Cassini/VIMS \citep{Maltagliati2015}, and ALMA  \citep{Molter2016}. We set constant vertical profiles for the constituents of our reference atmosphere above their respective condensation pressure level. In our model, the abundances of Titan's atmospheric constituents do not depend on the considered latitude. However, \citet{Lellouch2014} showed that in the lower stratosphere (around 15~mbar) the \methane~mole fraction varies significantly (from $1.0\%$ to $1.5\%$), which can affect the temperatures and gas abundances retrieved in this study.  We address this problem in section \ref{sect_err}. \\
 
                        Spectroscopic data for the constituents of the reference atmosphere come from the GEISA 2015 database \citep{Jacquinet-Husson2016}. The spectral contributions due to collisions induced absorption (CIA) between the main constituents of Titan's atmosphere ($\mathrm{N_2}$, \methane, and $\mathrm{H_2}$) are calculated using the studies of \citet{BorysowFrommhold1986,BorysowFrommhold1986a,Borysow1986,BorysowFrommhold1987}, \citet{BorysowTang1993}, and \cite{Borysow1991}. 
                        Following the studies of  \citet{Tomasko2008c,deKok2010}, we multiply the absorption coefficients of the CIA by 1.5.\\             
                         
\setcounter{table}{1}
                         
                        \begin{table}[h]
                                \caption{Constituents of the reference atmosphere, their volume mixing ratios, and the studies from which these measurements come from. $\mathrm{N_2}$ abundance has been normalised so that the total sum of the volume mixing ratios of all the gases of our reference atmosphere is equal to 1.  Asterisks denote the gases for which the abundances are retrieved. Their volume mixing ratios are {a priori} values.}
                                \label{tab_atm}
                                \centering
                                \begin{tabular}{ccc}
                                        \hline\hline
                                        Gas & Volume Mixing Ratio & References \\
                                        \hline
                                        $\mathrm{N_2}$ &  0.9839 & Normalisation \\
                                        \methane & 0.0148  &  \citet{Niemann2010} \\
                                        $\mathrm{^{13}CH_4}$ & $1.71 \times 10^{-4} $ &  \citet{Nixon2012} \\
                                        $\mathrm{CH_3D}$ & $9.4 \times 10^{-6} $ &  \citet{Nixon2012} \\
                                        $\mathrm{H_2O}$ & $1.4\times 10^{-10}$ & \citet{Cottini2012} \\
                                        $\mathrm{H_2}$ & $1.01\times 10^{-3}$ & \citet{Niemann2010} \\
                                        *$\mathrm{C_2N_2}$& $2.0\times 10^{-10}$ & \citet{Teanby2009}\\
                                        *$\mathrm{C_3H_4}$& $1.2\times 10^{-8}$ & \citet{Teanby2009}\\
                                        *$\mathrm{C_4H_2}$& $2.0\times 10^{-9}$ & \citet{Teanby2009} \\
                                        $\mathrm{CO_2}$& $1.6\times 10^{-8}$ & \citet{Coustenis2016} \\
                                        $\mathrm{HCN}$& $7.0\times 10^{-8}$ & \citet{Molter2016} \\
                                        $\mathrm{HC_3N}$& $5.0\times 10^{-10}$ & \citet{Coustenis2016} \\              
                                        $\mathrm{C_2H_2}$& $3.0\times 10^{-6}$ & \citet{Coustenis2016}\\
                                        $\mathrm{C_2H_4}$& $1.0\times 10^{-7}$ & \citet{Coustenis2016} \\
                                        $\mathrm{C_2H_6}$& $1.0\times 10^{-5}$ & \citet{Coustenis2016} \\
                                        $\mathrm{C_3H_8}$& $1.5\times 10^{-6}$ & \citet{Coustenis2016} \\
                                        $\mathrm{CO}$& $4.6\times 10^{-5}$ & \citet{Maltagliati2015} \\
                                        $\mathrm{C_6H_6}$& $4.0\times 10^{-10}$ & \citet{Coustenis2016} \\      
                                        \hline
                                        
                                \end{tabular}
                         \end{table}

                        Our reference atmosphere takes into account the broad spectral contributions of Titan's stratospheric hazes. For the FP1 spectra ($70-400~\mathrm{cm^{-1}}$), we consider four types of hazes, following \citet{deKok2007}.  The main feature (haze 0) covers all wavenumbers from $70~\mathrm{cm^{-1}}$ to $550~\mathrm{cm^{-1}}$. Its extinction profile has a scale height of 65~km from 80~km to 250~km and is constant below 80~km, following \citet{deKok2010,Tomasko2008}.   Three other localised features are also included: hazes A (centred at $140~\mathrm{cm^{-1}}$), B (centred at $220~\mathrm{cm^{-1}}$), C (centred at $190~\mathrm{cm^{-1}}$) as described in \citet{deKok2007}. For the FP4 spectra ($1200-1360~\mathrm{cm^{-1}}$), we use the aerosol properties measured by \citet{Vinatier2012}.   
                         
                        \subsection{Retrieval method}
                                \label{sect_retrieval}
                                The retrievals are performed in several iterations. To retrieve a given variable (e.g. temperature profile, scale factor toward a given {a priori} vertical profile of a gas) at each iteration, a synthetic spectrum is calculated by NEMESIS using the correlated-k method. Then, the difference between the synthetic and the measured spectra is used to compute an increment to add to the retrieved variable. At the next step, the new value or profile of this variable is then used to compute a new synthetic spectrum.  This method is detailed in \citet{Irwin2008}. \\

                                We retrieve simultaneously a continuous temperature profile, and best-fitting scales factors for the {a priori} vertical profiles of \cyano, \methyla, \diacety, hazes 0, A, B, and C from the FP1 spectra. Hazes fit the continuum component of the spectra. Temperature is measured using the radiance in the ten \methane~rotational bands between $70~\mathrm{cm^{-1}}$ and $170~\mathrm{cm^{-1}}$, and the continuum. Abundances of \diacety, \cyano, and \methyla~are obtained by fitting the radiance in their respective spectral bands at $220~\mathrm{cm^{-1}}$, $234~\mathrm{cm^{-1}}$, and $327~\mathrm{cm^{-1}}$.\\
                        
\begin{figure}[!h]
                                        \begin{center}
                                        \includegraphics[width=1\columnwidth]{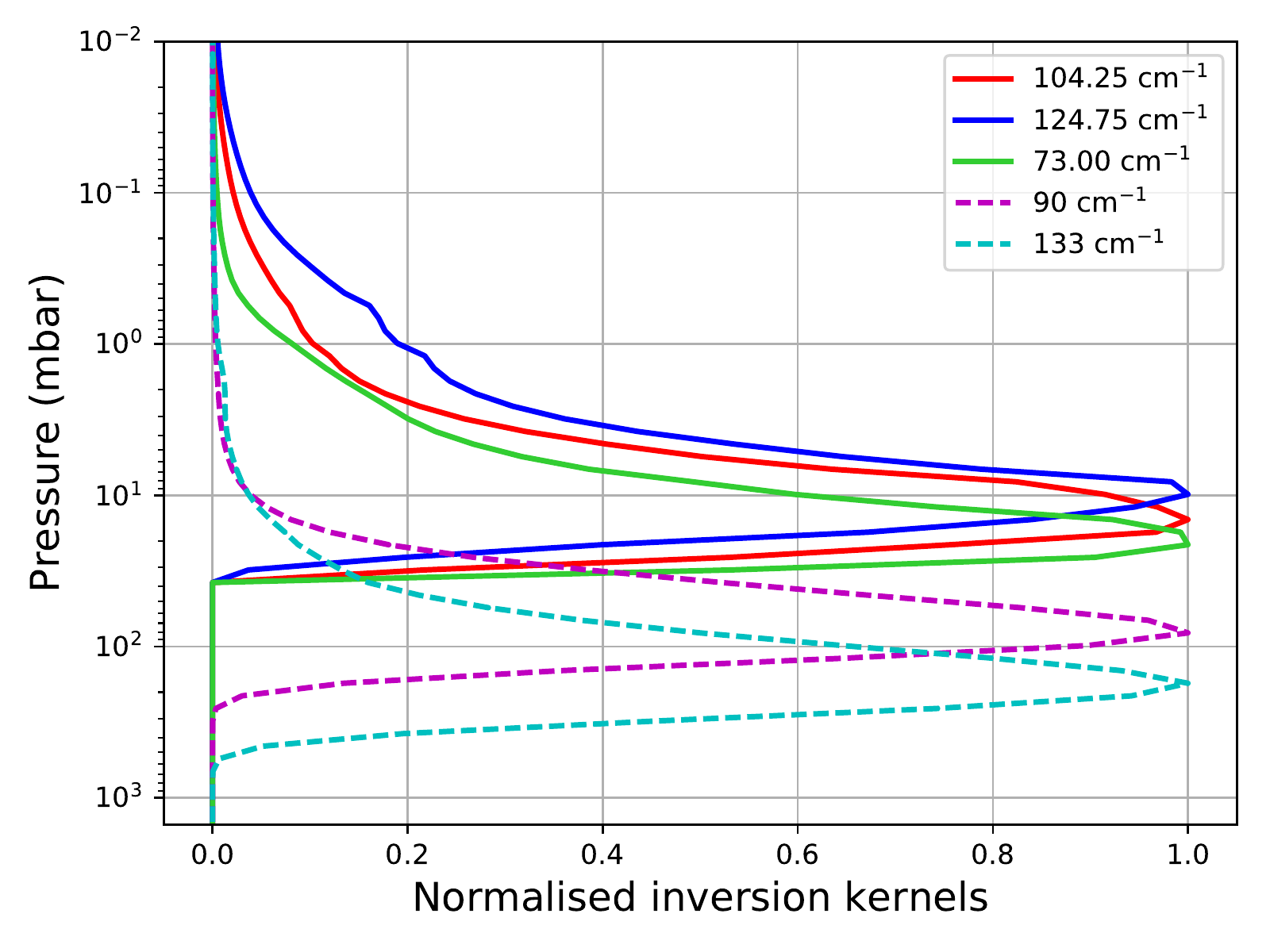}
                                        \caption{Normalised inversion kernels for temperature retrievals from the FP1 spectra. Solid lines are inversion kernels obtained within three rotational \methane~bands. Dashed lines are inversion kernels for  two wavenumbers in the continuum. \methane~rotational bands (from $70~\mathrm{cm^{-1}}$ to $170~\mathrm{cm^{-1}}$) probe the stratospheric temperature between 10~mbar and 20~mbar. Wavenumbers in the continuum probe the temperature around the tropopause from 80~mbar to 200~mbar.}
                                        \label{fig_fcontrib_temp_fp1}
                                        \end{center}
                                
                                \end{figure}

\begin{figure}[!h]
                                        \begin{center}
                                                \includegraphics[width=1\columnwidth]{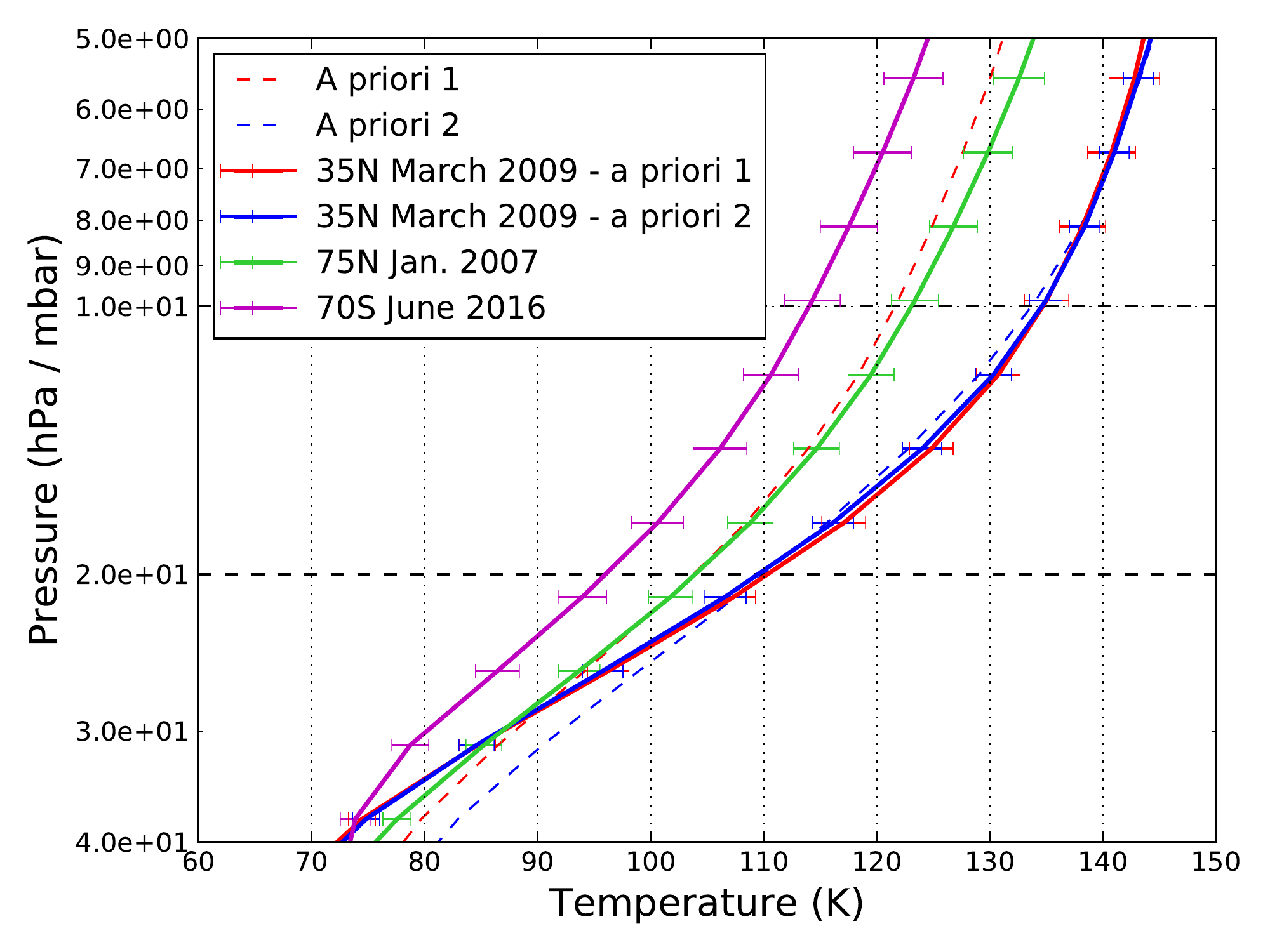}
                                                \caption{Examples of temperature profiles over the whole range of  temperature profiles retrieved in this study. For the observations at $35^{\circ}$N in March 2009, we show temperature profiles obtained with two different {a priori profiles}. {A priori} 1 is a temperature profile measured by \citet{Achterberg2008}, while {a priori} 2 is the profile retrieved from FP4 observations performed at $30^{\circ}$N in June 2009.  Black horizontal dashed lines show the sensitivity limits of the FP1 temperature retrievals. Error bars on the profiles do not take into account the errors related to \methane~variations (see section \ref{sect_err}).}
                                                \label{fig_comp_ap}
                                \end{center}
                         \end{figure}

                                For each retrieved physical quantity, we can assess the sensitivity of our measurements as a function of the pressure using the inversion kernels defined as  
                        
                                \begin{equation}
                                        K_{ij} = \frac{\partial I_i}{\partial x_j}
                                ,\end{equation}
                        
                                \noindent where $I_i$ is the measured radiance at the wavenumber $w_i$, and $x_j$ the value of a given retrieved parameter (e.g. temperature, scale factor toward the {a priori} profile for a gas) at the pressure level $p_j$. Figure \ref{fig_fcontrib_temp_fp1} shows the inversion kernels for temperature at wavenumbers within the continuum ($90~\mathrm{cm^{-1}}$ and $133~\mathrm{cm^{-1}}$)  and three rotational \methane~bands ($73~\mathrm{cm^{-1}}$, $104.25~\mathrm{cm^{-1}}$, and $124.75~\mathrm{cm^{-1}}$). The continuum emission depends on the  extinction profile of haze 0, and on the temperature near the tropopause (between 80~mbar and 200~mbar). The \methane~bands allow us to measure the temperature in the lower stratosphere between 10~mbar and 20~mbar. Figure \ref{fig_comp_ap} shows representative examples of retrieved temperature profiles in the region probed by the \methane~bands.\\
                
\begin{figure}[!h]
                                        \begin{center}
                                                \includegraphics[width=1\columnwidth]{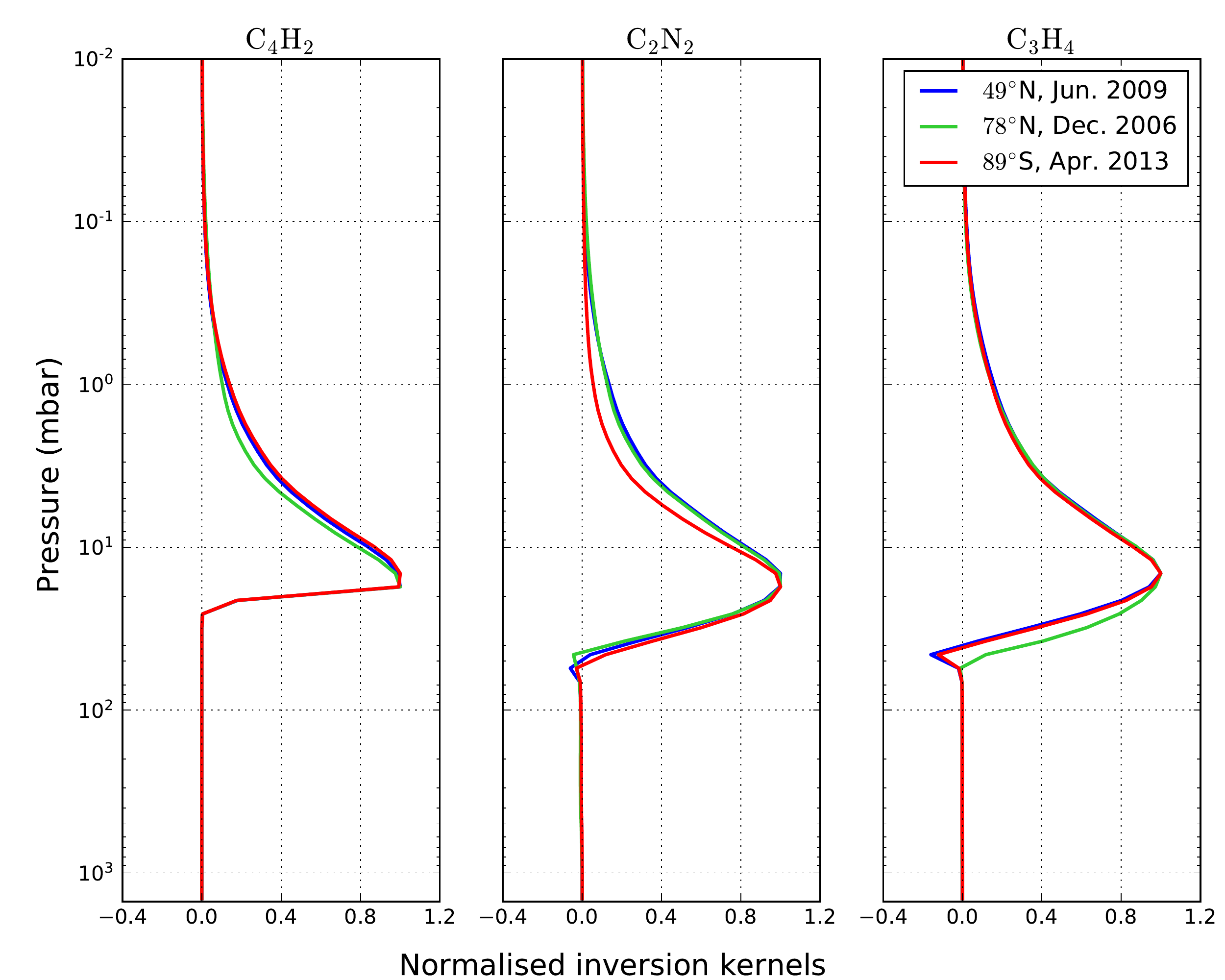}
                                                \caption{Normalised inversion kernels for \diacety~($220.25~\mathrm{cm^{-1}}$), \cyano~($234~\mathrm{cm^{-1}}$), and \methyla~($326.75~\mathrm{cm^{-1}}$) at different latitudes. The FP1 data probe the lower stratosphere between 5~mbar and 15~mbar; the  peak sensitivity is at 15~mbar.}
                                                \label{fig_wf_lat}
                                        \end{center}
                                \end{figure}

                                Figure \ref{fig_wf_lat} shows the normalised inversion kernels of \diacety, \cyano, and \methyla\ plotted respectively at $220.25~\mathrm{cm^{-1}}$, $234~\mathrm{cm^{-1}}$, and $326.75~\mathrm{cm^{-1}}$ for different latitudes and times. For the three species, at all latitudes, the maxima of the inversion kernels are at 15~mbar (85~km). This is deeper than the average pressure levels probed by Cassini/CIRS mid-infrared limb (from 5~mbar to 0.001~mbar) and nadir observations ($\sim10$~mbar). The width of the contribution function varies slightly throughout the different datasets, but the pressure level of the maximum stays constant.  \\   
                        
                                To  evaluate the robustness of our results, we perform retrievals with a wide range of {a priori} temperatures and compositions. We use different  {a priori} temperature profiles from \citet{Achterberg2008} to retrieve the temperature from the FP1 spectra.  When there are FP4 spectra acquired at the same latitude  (within $\pm 5^{\circ}$) and time (within 3 months) as the considered FP1 spectra, we retrieve a temperature profile from the FP4 spectra and use it as an  {a priori} in an additional FP1 retrieval.   Indeed, the \methane~$\nu_4$~band ($1304~\mathrm{cm^{-1}}$) visible in the FP4 spectra is sensitive to the temperature between 2~mbar and 0.5~mbar, with a peak sensitivity at 1~mbar. Figure \ref{fig_comp_ap} shows examples of temperature retrievals performed on the same dataset  (FP1 spectra measured at $35^{\circ}$N in March 2009), with different  {a priori} profiles.  Between 10 ~mbar and 20~mbar, the temperature profiles retrieved from FP1 do not depend on the chosen  {a priori} profile. Above the 10~mbar pressure level, the retrieved temperature profiles tend toward their respective  {a priori}, whereas below the 20~mbar level,  they are influenced by the information coming from the continuum emission. We also use several scale factors toward the  {a priori} profiles of hazes and retrieved gases for FP1, and various errors on these profiles or scale factors. These tests show that our results are not sensitive to  {a priori} assumptions, and allow us to find the best fit of the measured spectra.\\ 
                        
                \subsection{Correction of error bars}
                        \label{sect_corr_err}
                        
                        \begin{figure}[!h]
                                \begin{center}
                                        \includegraphics[width=1\columnwidth]{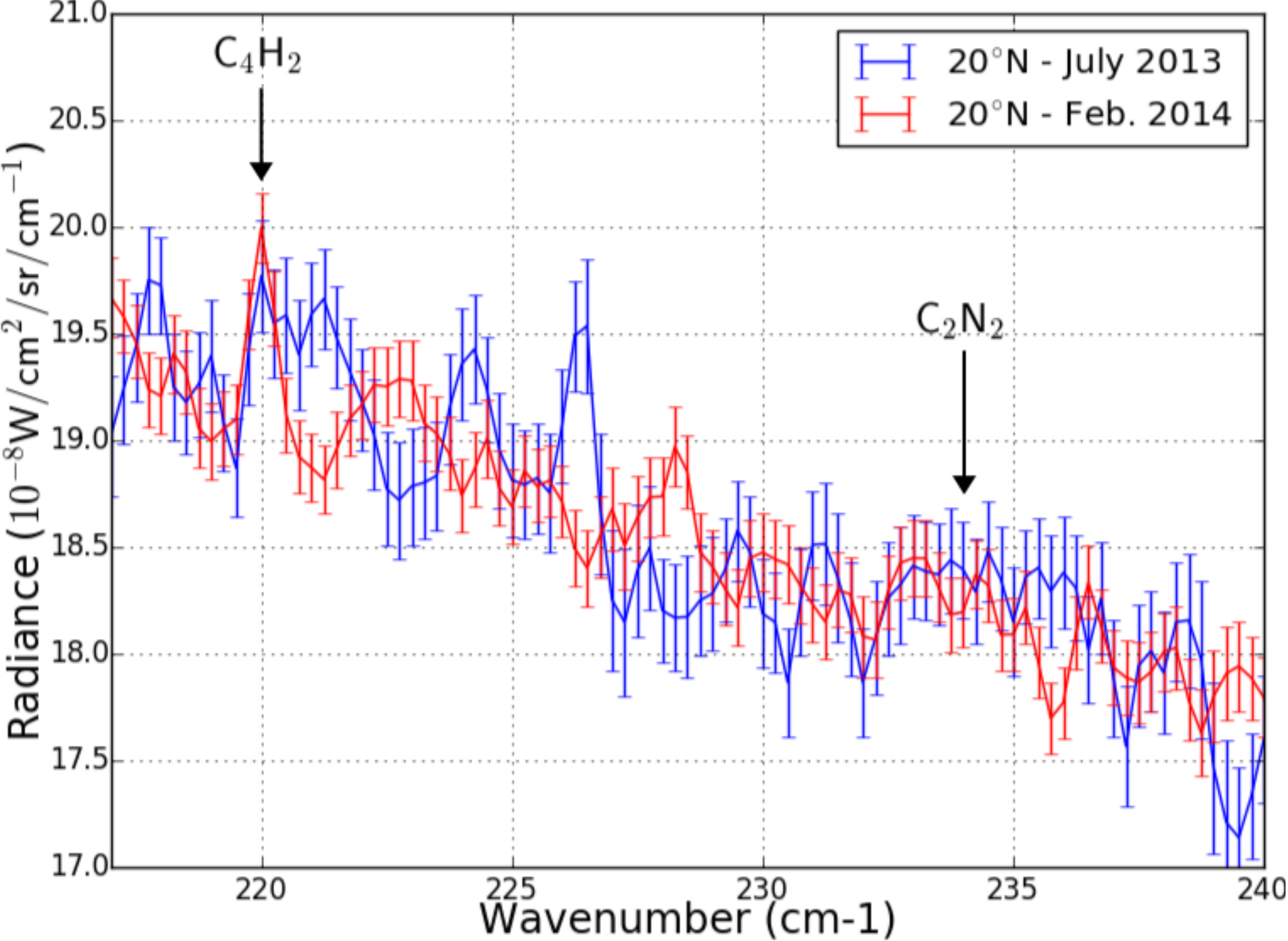}
                                        \caption{Spectra measured at $20^{\circ}$N in July 2013 (blue) and February 2014 (red), between $217~\mathrm{cm^{-1}}$ and $240~\mathrm{cm^{-1}}$. Here we show the error bars from the photometric calibration provided by the CIRS team. These error bars are too small to take into account the spurious noise features in each spectrum or some of the radiance differences between these two spectra.}
                                        \label{fig_bruit}
                                \end{center}    
                                
                        \end{figure}
                        
                        We noticed that in the FP1 spectra, error bars provided by the calibration are too small, i.e. they do not take into account all the noise 
                        radiance variations of the spectra. Figure \ref{fig_bruit} shows two averaged spectra obtained at $20^{\circ}$N at two close dates (July 2013 and February 2014). We selected two datasets where the signal-to-noise ratio is particularly low as it makes the issue described here more visible. In this example, the average radiances of these spectra are equal, but the radiance variations due to the noise features are too large with respect to the error bars. As this can lead to an overestimation of the level of detection of the studied species, we make our own estimation of the error bars by measuring the radiance variations due to noise around the spectral bands of \diacety~(220$~\mathrm{cm^{-1}}$), \cyano~($234~\mathrm{cm^{-1}}$), \methyla~($327~\mathrm{cm^{-1}}$).\\
                        
                        To compute the new error bars, for each FP1 spectrum, we first fit the continuum component, i.e. the spectrum without \diacety~(219-222$~\mathrm{cm^{-1}}$), \cyano~(233-236$~\mathrm{cm^{-1}}$), and \methyla~(322-334$~\mathrm{cm^{-1}}$). Then, we define the following domains in regions of the spectra without strong emission lines around the retrieved gases: 
                
                        \begin{itemize}
                                \item $[214; 219[ \cup ]222; 225]~\mathrm{cm^{-1}}$ (around the \diacety~band)
                                \item $[228.5; 233[ \cup ]236; 241]~\mathrm{cm^{-1}}$ (around the \cyano~band)
                                \item $[317; 322[ \cup ]334; 337]~\mathrm{cm^{-1}}$ (around the \methyla~band)
                        \end{itemize}
                        
                        In each of these spectral domains, we  estimate the new error bars $\sigma$ with  
                        
                        $$\sigma = \max{(I_{mes} - I_{cont})}$$ 
                        
                        \noindent with $I_{mes}$ the measured radiance and $I_{cont}$ the synthetic spectrum of the continuum. We use $\sigma$ as the new value of the minimum error in the whole domain (including the band of the considered gas).  

                \subsection{Correction of the continuum at high latitudes during autumn and winter}
                        \label{sect_conti}
                        
                        \begin{figure}[h]
                                \centering
                                \includegraphics[width=1\columnwidth]{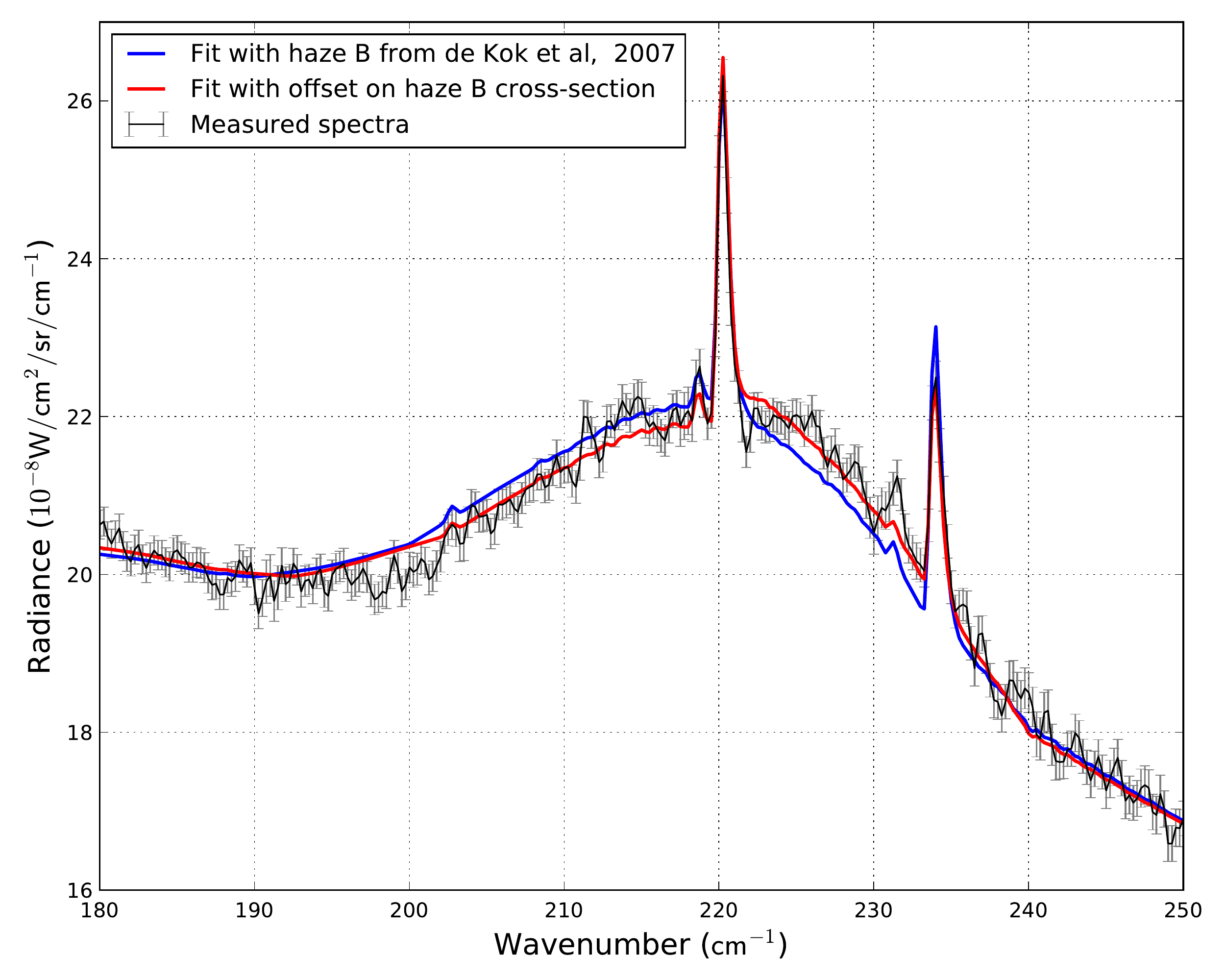}
                                \caption{Fits of a high latitude spectrum during winter.  The black spectrum was measured with FP1 at $89^{\circ}$N in March 2007. The blue and red lines indicate respectively the  fits of this spectrum with the haze B cross-sections as measured by \citet{deKok2007} and with the addition of a small offset ($2.5~\mathrm{cm^{-1}}$) to haze B cross-sections from $190~\mathrm{cm^{-1}}$ to $240~\mathrm{cm^{-1}}$. Without this correction, the continuum is overestimated by NEMESIS around the \diacety~band and underestimated around the \cyano~band.}
                                \label{fig_conti}
                        \end{figure}

                        We note that in several spectra acquired at high northern and southern latitudes (from $70^{\circ}$N/S to $90^{\circ}$N/S) during their respective winter (2006-2007) and autumn (2014-2016), the continuum component has a different shape in the $190 - 240~\mathrm{cm^{-1}}$ range from that in the other spectra acquired at different latitudes or seasons. This shape is characterised by a broad emission feature centred at $220~\mathrm{cm^{-1}}$. Previous studies such as \citet{Coustenis1999,Anderson2012,Jennings2012,Jennings2015} measured the same spectral feature in Voyager/IRIS and Cassini/CIRS data, at similar latitudes and seasons, and suggested that it could be caused by a mixture of nitrile condensates. This new shape of the continuum could not be fitted correctly with temperature, and the cross-sections and vertical distributions measured for hazes 0 and B by \citet{deKok2007,deKok2010}. For instance, in figure \ref{fig_conti}, we compare the spectrum measured at $89^{\circ}$N in March 2007 and its fit by NEMESIS using our nominal parameters for the hazes. Between $190~\mathrm{cm^{-1}}$ and $222~\mathrm{cm^{-1}}$, the radiance of the continuum of the measured spectrum is lower than the radiance of the continuum fitted by Nemesis, whereas there is the opposite situation from $222~\mathrm{cm^{-1}}$ to $240~\mathrm{cm^{-1}}$. Wavenumbers from $240~\mathrm{cm^{-1}}$ to $400~\mathrm{cm^{-1}}$ are not affected by this feature. This change in the continuum shape is an issue when trying to fit \diacety, \cyano, and \methyla, as it affects the fits of \diacety~and \cyano~bands, and leads to an underestimation of the abundance of \diacety~and an overestimation of the abundance of \cyano.\\
                        
                        However, we note that the continuum in the haystack region ($190 - 240~\mathrm{cm^{-1}}$) can be fitted by adding a small offset in wavenumber to the cross-sections of haze B for the affected datasets.  For the spectra measured at high northern latitudes during the winter (2006-2007), this offset is  between $2~\mathrm{cm^{-1}}$ and $2.5~\mathrm{cm^{-1}}$, while its value is between $2.5~\mathrm{cm^{-1}}$ and $3~\mathrm{cm^{-1}}$ for the observations of high southern latitudes during autumn (2012-2016). This offset would be consistent with the appearance of new condensates as suggested by the previous studies. The small difference between the offset values in the northern and southern high latitudes may be due to the fact that they were observed at different seasons (northern winter and southern autumn), and thus different stages of the chemical evolution at the poles. 
                        
                \subsection{Upper limits}               
                
                        For each FP1 dataset, we evaluate the level of detection of \diacety, \cyano, and \methyla. For each gas, we define the $\chi^2$ as  
                        
                        \begin{equation}
                                \chi^2 = \sum_{i=1}^{N} \frac{\left(I_{mes}(w_i) - I_{fit}(w_i, x)\right)^2}{2\sigma_i^2}
                        ,\end{equation}  
                                
                        \noindent where $I_{mes}(w_i)$ and $I_{fit}(w_i, x)$ are respectively the radiance measured at the wavenumber $w_i$ and the fitted radiance at the same wavenumber for the volume mixing ratio $x$ of the considered gas, $N$ is the total number of points in the measured spectra, and   $\sigma_i$ is the error on the radiance measured at the wavenumber $w_i$.  The factor 2 in the denominator is the oversampling factor of the data.\\ 
                        
                        \noindent Then we compute the misfit $\Delta \chi^2$ defined as 
                        \begin{equation}
                                \Delta\chi^2 = \chi^2 - \chi_{0}^2 
                        ,\end{equation} 
                        \noindent where $\chi_0^2$ is the $\chi^2$ obtained when we fix the abundance of the considered gas to $x=0$. The 1-$\sigma$, 2-$\sigma$, or 3-$\sigma$ detection level is reached when $\Delta\chi^2$ is respectively inferior to -1, -4, or -9. Most of the spectra acquired at high and mid-northern latitudes and at the south pole in autumn allow us to measure the abundances of \cyano, \methyla, and \diacety~with a confidence level greater than 3-$\sigma$.\\
                        
                        For some datasets, mostly the observations in the equatorial region, at southern mid-latitudes, and at southern high latitudes during summer, the signal-to-noise ratio is not good enough to detect the spectral bands of one or several of the studied species. For these datasets, for each undetected gas, we obtain an upper limit of its volume mixing ratio $x$ by calculating synthetic spectra for different values of $x$, starting with $x=0$ and incrementing it progressively. For each of these synthetic spectra, we compute the misfit $\Delta\chi^2$. The value of $x$ for which $\Delta\chi^2$ is minimum is the upper limit for the volume mixing ratio of the considered gas. In these cases, $\Delta\chi^2$ is positive and values of 1, 4, and 9 respectively indicate a 1-, 2-, or 3-$\sigma$ upper limit.    
                
                \subsection{Error analysis}     
                        \label{sect_err}
                        Our retrievals are mainly affected by three error sources: measurement noise, errors related to the retrieval process (e.g. smoothing of the retrieved profile, forward modelling error), and the uncertainty on the \methane~abundance.  The effects of the first two error sources are directly estimated by NEMESIS. \citet{Lellouch2014} showed that at 15~mbar, \methane~abundance varies from 1.0\% to 1.5\%. The upper value is consistent with the measurements from \citet{Niemann2010}, which is the \methane~abundance used in our reference atmosphere. We perform additional retrievals to evaluate how a \methane~abundance as low as 1.0\% would affect our results. We find that the temperature retrieved at 15~mbar would increase by 4-5~K (which is consistent with the results of \citet{Lellouch2014}), and that the uncertainty on the \methane~abundance is the dominant error source for the temperature retrievals. This temperature change would also decrease the retrieved volume mixing ratios of \cyano, \methyla, and \diacety. For \cyano~and \methyla, the difference between retrievals performed with $1.0\%$ and $1.48\%$ of \methane~is comparable to the combined effect of measurement noise and retrieval errors. For \diacety, the effect of \methane~variation is twice as big as that of the other error sources. These results are summarised in table \ref{tab_error}.
                 
                        \begin{table*}[!h]
                                \caption{Error estimation on the temperature and gases retrievals.}
                                \label{tab_error}
                                \centering
                                \bgroup
                                \def\arraystretch{1.5}
                                \begin{tabular}{ccccc}
                                        \hline
                                        \hline
                                                                 &             Noise and retrieval               &   $[\mathrm{CH_4}]$~variations &       Quadratic sum   \\
                                        \hline
                                        Temperature & $\pm 1.6\%$                               &    $+3\%$       &          $\substack{+3.4\%\\-1.6\%}$  \\
                                        \hline
                                        \cyano                  & $\pm 16\%$                            &         $-16\% $ &     $\substack{+16\%\\-23\%}$  \\
                                        \hline
                                        \methyla                        & $\pm 7\%$                               &       $-5\% $ &     $\substack{+7\%\\-9\%}$  \\
                                        \hline
                                        \diacety                        & $\pm 7\%$                               &       $-15\% $ &     $\substack{+7\%\\-17\%}$  \\
                                        \hline
                                \end{tabular}
                        \egroup
                \end{table*}

\section{Results}       
        \label{sect_res}
        \subsection{Radiance evolution at high latitudes}
                \begin{figure}[!h]
                        \begin{center}
                                \includegraphics[width=1\columnwidth]{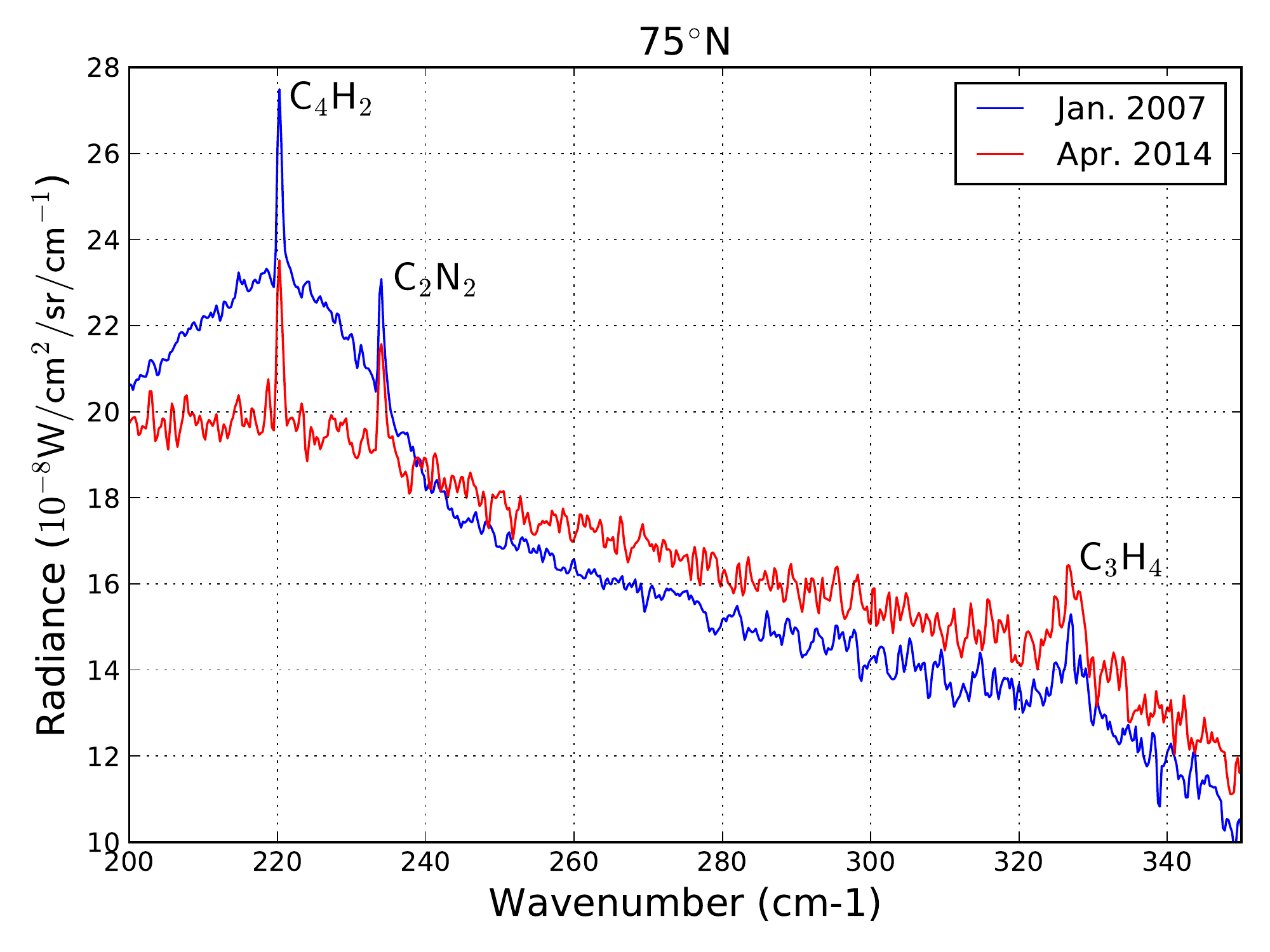}

                                \includegraphics[width=1\columnwidth]{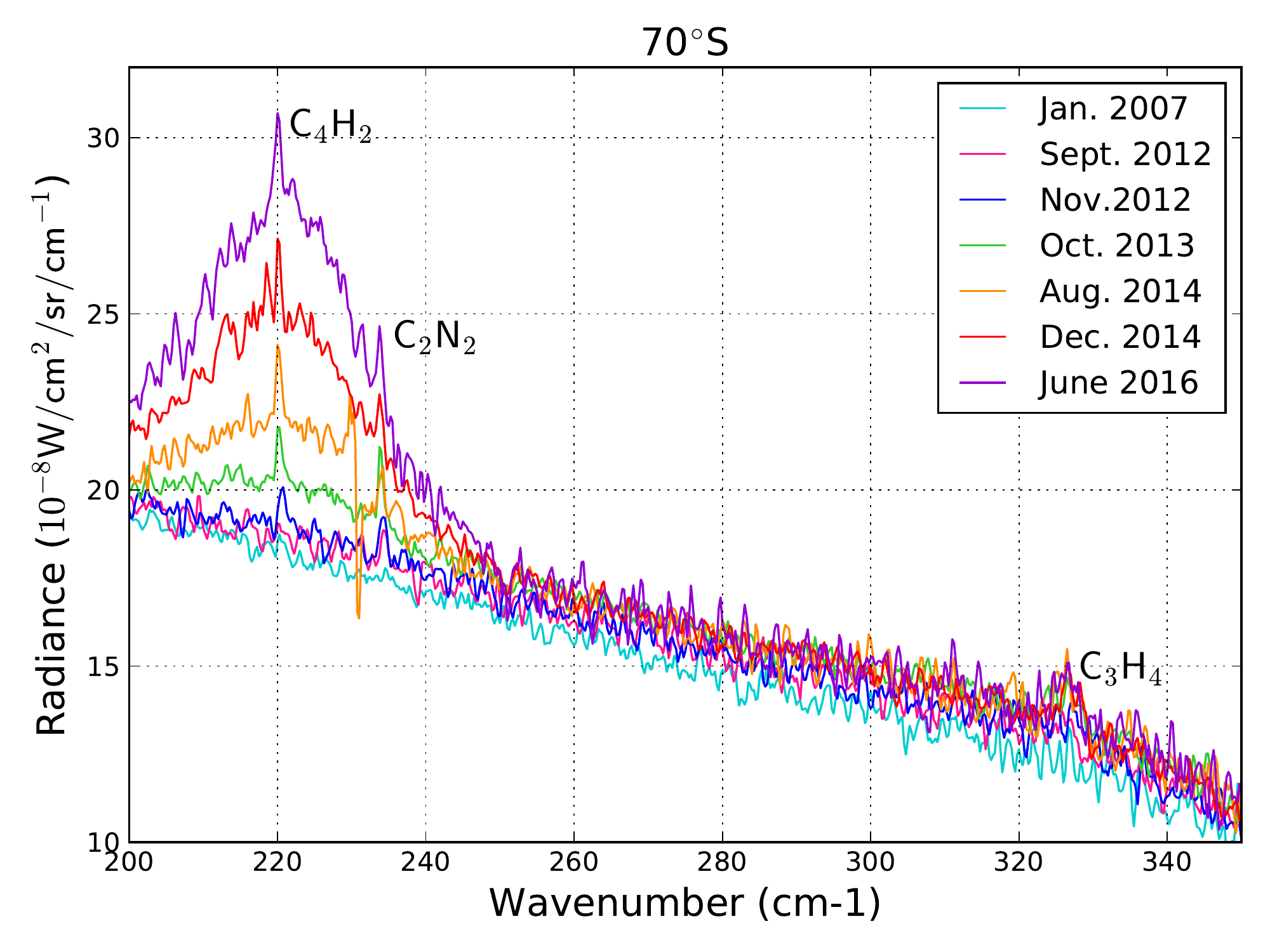}
                                \caption{Evolution of measured radiances at $75^{\circ}$N (top panel) and $70^{\circ}$S (bottom panel) from 2007 (northern winter) to 2016 (mid-spring). The sharp variation in the radiance between $228~\mathrm{cm^{-1}}$ and $231~\mathrm{cm^{-1}}$ in the spectra measured in August 2014 is a spurious noise feature which could not be eliminated during the calibration process. From 2007 to 2016, radiance at high southern latitudes has strongly increased, whereas it has decreased at high northern latitudes.}
                                \label{fig_ev70NS}
                        \end{center}    
                \end{figure}
        
                Figure \ref{fig_ev70NS} shows the spectra measured at $75^{\circ}$N (top panel) in 2007 and 2014, and at  $70^{\circ}$S between 2007 and 2016. At both latitudes there is a striking evolution of the measured radiances as Titan's atmosphere goes from northern winter to spring. Radiance variations of this amplitude are only observed at high northern and southern latitudes. In both cases presented in fig. \ref{fig_ev70NS}, the largest change in the measured radiance occurs between $200~\mathrm{cm^{-1}}$ and $250~\mathrm{cm^{-1}}$. At $75^{\circ}$N the broad emission feature centred on $220~\mathrm{cm^{-1}}$ present during northern winter (2007) completely disappeared in mid-spring. The amplitude of the spectral bands of \diacety~($220~\mathrm{cm^{-1}}$) and \cyano~($234~\mathrm{cm^{-1}}$) slightly decreased from 2007 to 2014, but these bands are still clearly visible in 2014.\\
                
                At $70^{\circ}$S, the temporal coverage of the data is better than at $75^{\circ}$N and allows us to follow more precisely the evolution of the radiance at this latitude. A broad emission feature centred at $220~\mathrm{cm^{-1}}$, similar to what was observed at the northern high latitudes appeared in 2013 and its radiance increased steeply from October 2013 to June 2016, while radiances between $250~\mathrm{cm^{-1}}$ and $400~\mathrm{cm^{-1}}$ (except in \methyla~band at $327~\mathrm{cm^{-1}}$) stayed constant during this period. The radiance in the broad emission feature (except in \diacety~band) is higher at $70^{\circ}$S in 2014 and 2016 (southern autumn) than at $75^{\circ}$N in 2007 (northern winter). The radiance in the bands of \diacety~($220~\mathrm{cm^{-1}}$) and \cyano~($234~\mathrm{cm^{-1}}$) also evolved rapidly from southern summer to mid-autumn. In 2007, the signal-to-noise ratio in the vicinity of the bands of \diacety\ and \cyano~was too low to distinguish these spectral features unambiguously. After September 2012, their radiances increased remarkably within a few months. The radiance of the \methyla~($327~\mathrm{cm^{-1}}$) band also increased during the same period, but more slowly than the other gases. The evolution of the radiances in the spectral bands of \diacety, \methyla, and \cyano~is faster at $70^{\circ}$S than at $75^{\circ}$N. The changes in the radiances at $75^{\circ}$N and $70^{\circ}$S suggest a significant seasonal evolution of the lower stratospheric composition.\\

        \subsection{Evolution of the meridional distributions of \diacety, \cyano, and \methyla}                   
                \begin{sidewaysfigure*}[!hp]
                        \begin{center}
                                \includegraphics[width=0.4\columnwidth]{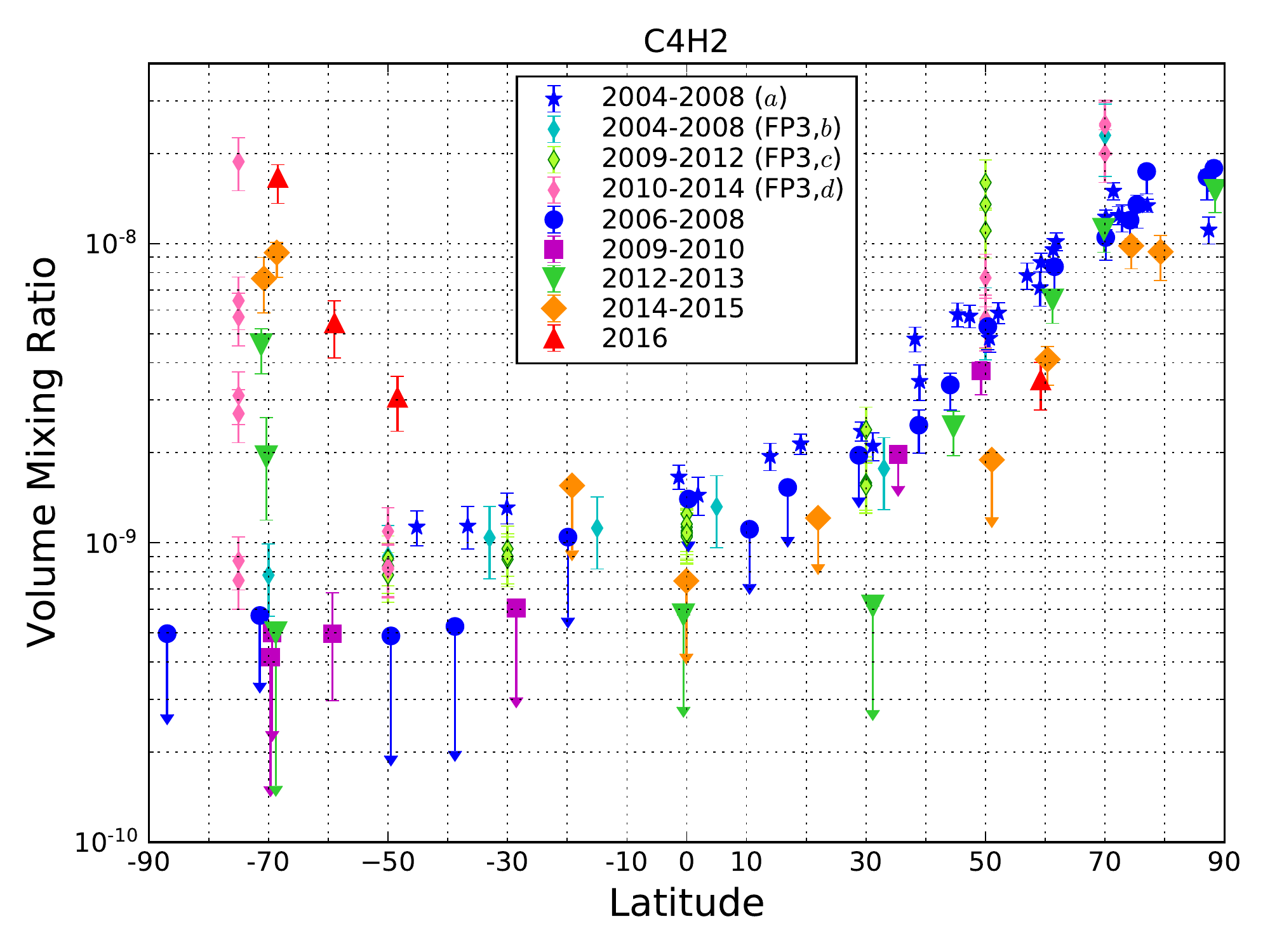}
                                \includegraphics[width=0.4\columnwidth]{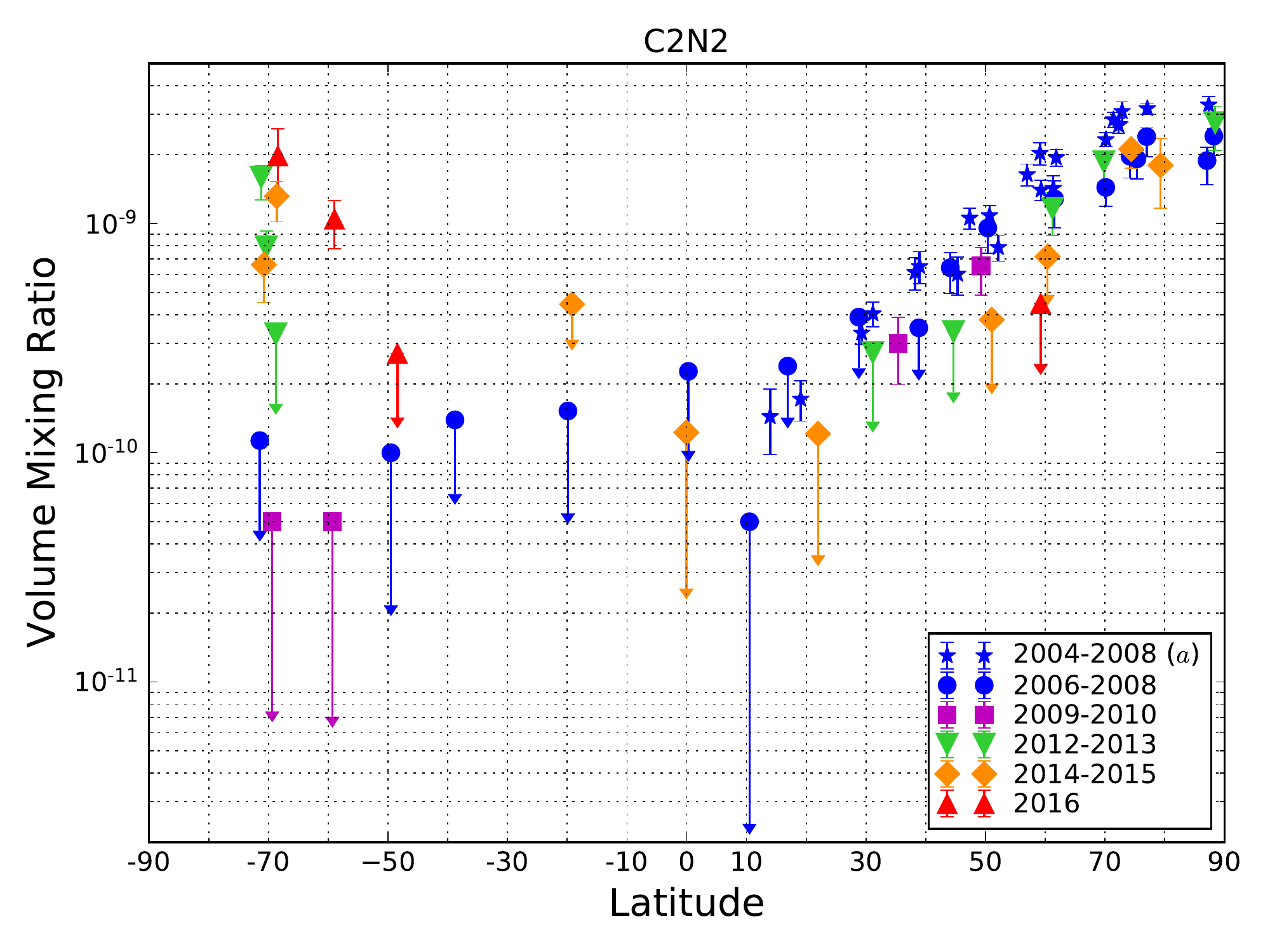}

                                \includegraphics[width=0.4\columnwidth]{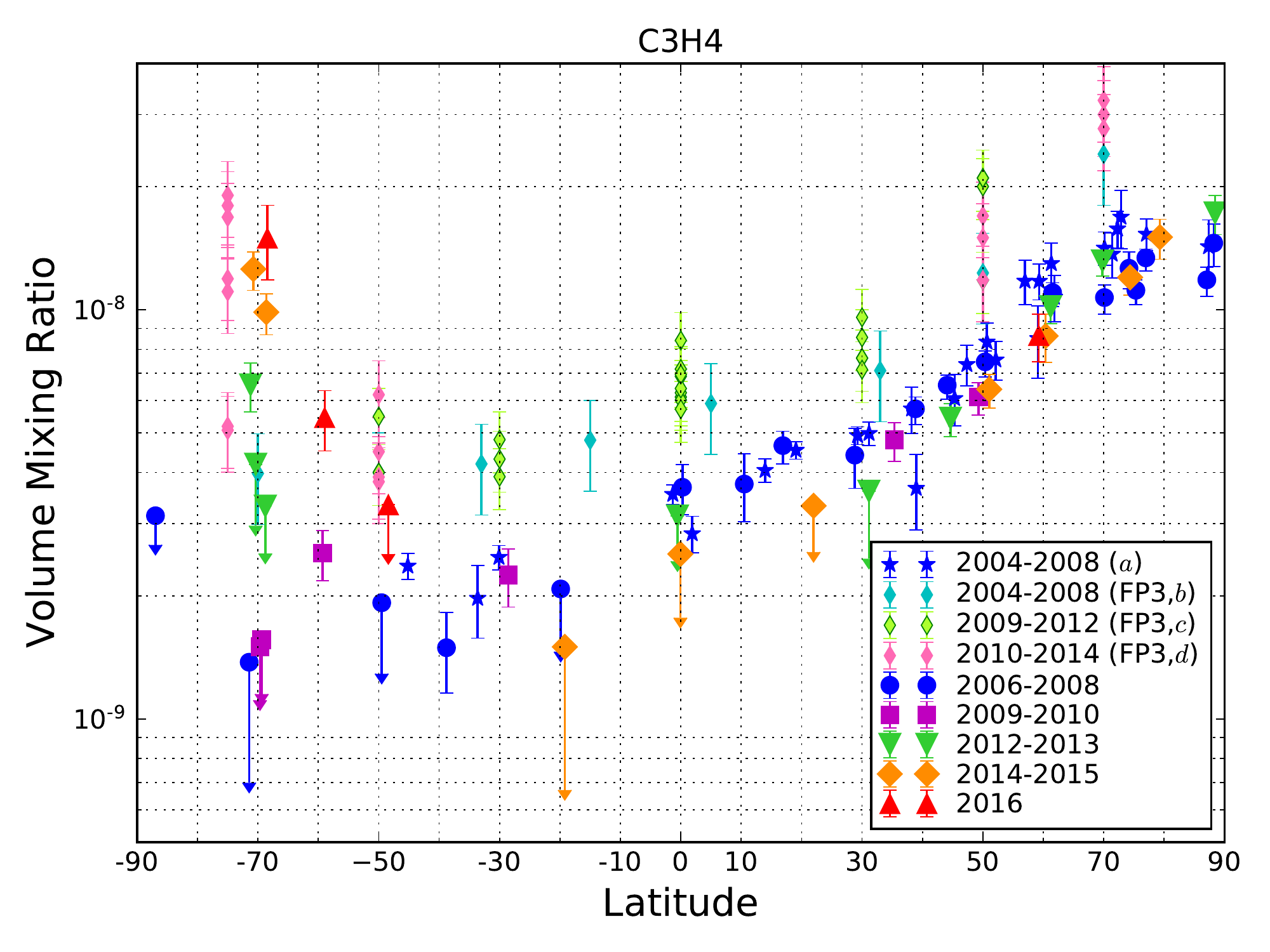}               
                                \caption{Meridional distributions of \diacety, \cyano, and \methyla~from 2006 (northern winter) to 2014 (mid-northern spring) at 15~mbar (or an altitude of $\sim$85~km). \textit{(a)} refers to  \citet{Teanby2009}, where the same gases as in this study were measured with the same CIRS detector (FP1);  \textit{(b)}, \textit{(c)}, and \textit{(d)} respectively refer to \citet{Coustenis2010}, \citet{Bampasidis2012}, and \citet{Coustenis2016},  \diacety~and \methyla~were measured with the CIRS detector FP3, probing slightly higher pressure levels (10~mbar) than this study. In the south pole, from 2006 to 2015, the volume mixing ratios of the three species  strongly increased, while the other latitudes exhibit weak seasonal variations. Unlike \diacety~and \methyla, \cyano~abundance at mid-northern latitudes decreases from 2012.  Error bars show relative errors as derived in section \ref{sect_err}.}
                                \label{fig_evmerid}                     
                        \end{center}
        \end{sidewaysfigure*}

                Figure \ref{fig_evmerid} shows the retrieved evolution of the meridional distributions of \diacety, \cyano, and \methyla~from 2006 (northern winter) to 2016 (late spring) at the 15~mbar pressure level (or an altitude of $\sim85~$km).\\
        
                The most striking feature in the plots of fig. \ref{fig_evmerid} is the sudden and steep increase of the abundances of \diacety, \cyano, and \methyla~from 2006 to 2016 at high southern latitudes (poleward of $70^{\circ}$S). During this period, at $70^{\circ}$S, abundances  increased by at least a factor of 39 for \diacety, 39 for \cyano, and 10 for \methyla. Most of this increase happened between 2012 and 2013. For instance, a factor of 2.4 can be measured between two consecutive measurements of \diacety~volume mixing ratios at $70^{\circ}$S during this period.\\      
        
                In contrast, the other latitudes show  smaller variations in the abundances of the studied gases. At high northern latitudes (poleward of $70^{\circ}$N), the volume mixing ratios of \diacety, \cyano, and \methyla~have stayed constant from 2006 to 2015. In the northern hemisphere, between $30^{\circ}$N and $70^{\circ}$N, \methyla~exhibits a different seasonal evolution from \diacety~and \cyano. Indeed, \methyla~abundance is constant from winter to late spring (from 2006 to 2016), whereas abundances of \diacety~and \cyano~are constant from winter to early spring (from 2006 to 2010), and then decrease in the middle of spring (after 2012).  In equatorial and mid-southern latitudes (from $25^{\circ}$N to $65^{\circ}$S), \diacety~and \methyla volume mixing ratios do not vary significantly from 2006 to 2015, then they increase in 2016 around $50^{\circ}$S-$60^{\circ}$S.  This evolution is sharp for \diacety~(increasing by a factor 11) and weaker for \methyla~(increasing by a factor 2). For \cyano, there are fewer data points at equatorial and mid-southern latitudes because of the weaker signal-to-noise ratio in this band, but it seems to follow the same evolution as \methyla~and \diacety.\\ 
                
                {The meridional distributions of the three gases follow the same trend in northern winter (2006-2008) and in early spring (2009-2010), with a decrease from the north pole to the south pole. This shape starts to evolve in mid-spring (2012) with the sudden enrichment in gases of the high southern latitudes. Then in late spring (2015-2016), the distributions of \cyano, \methyla, and \diacety~at other latitudes slowly begin to evolve toward a more symmetrical shape, with a decrease from poles to equator.\\

\section{Discussion}
\label{sect_discu}

        \subsection{Comparison with previous Cassini/CIRS measurements}
                 In fig. \ref{fig_evmerid}, our measurements are compared to the results from previous Cassini/CIRS observations of Titan's lower stratosphere. These abundances were inferred from  
                \begin{itemize}
                        \item nadir FP1 $0.5~\mathrm{cm^{-1}}$ resolution spectra  (same type of data as in this study) from \citet{Teanby2009} during northern winter (2004-2008);
                        \item nadir FP3 $0.5~\mathrm{cm^{-1}}$ resolution spectra from \citet{Coustenis2010} during northern winter (2004-2008), \citet{Bampasidis2012} during early northern spring (2009-2012), and \citet{Coustenis2016} during mid-northern spring (2010-2014).   
                \end{itemize}

                Our results for the period 2006-2008 are  in overall good agreement with the results of \citet{Teanby2009}. The \cyano~and \methyla~abundances measured in this study from 2006 to 2008 are similar to the values of \citet{Teanby2009}. We find slightly lower \diacety~abundances than they do, but this is probably due to the update of the spectroscopic parameters for the \diacety~band at $220~\mathrm{cm^{-1}}$ in GEISA 2015.\\
        
                The comparison between our results for \methyla~and \diacety~and the FP3 measurements of \citet{Coustenis2010,Coustenis2016} and \citet{Bampasidis2012} shows that although we obtain values of the same order of magnitude, we often measure lower abundances. This is particularly visible on \methyla ~(at all seasons and latitudes), whereas \diacety~abundances inferred in this study are  similar to \citet{Coustenis2010}  in 2006-2008, then lower than  \citet{Bampasidis2012} and \citet{Coustenis2016} after 2009. As these studies were performed in the  $600-1100~\mathrm{cm^{-1}}$ region, \diacety~and \methyla~were not probed with the same spectral bands as those in our study. This might be responsible for the small disparities between the results from nadir FP3 observations and this study. Differences between the retrieval codes (NEMESIS and ARTT) may also be at play.  FP1 and FP3 results may also be different because FP3 nadir observations probe slightly lower pressures (around 10 mbar or $\sim100~$km) than FP1 nadir observations (around 15 mbar or $\sim85~$km). This would be consistent with the predictions of photochemical models such as \citet{Wilson2004}, \citet{Krasnopolsky2014} and \citet{Dobrijevic2016}, where \methyla~and \diacety~profiles increase weakly with altitude in Titan's lower stratosphere. \\

        \subsection{Interannual variations}

                In table \ref{tab_comp_Voyager_Cass}, we compare the results inferred from Voyager I/IRIS observations by \citet{Coustenis1995} with the abundances of \diacety, \methyla, and \cyano~measured with Cassini/CIRS in this study. While we measure similar abundances at $30^{\circ}$N, we obtain values slightly lower at  $30^{\circ}$S,  and significantly lower at $50^{\circ}$N than \citet{Coustenis1995}.  The largest difference between these two sets of measurements is reached for \cyano~ at $50^{\circ}$N:  there is a factor of 23 difference between this study and the Voyager results. \citet{Coustenis2013} compared Voyager and Cassini/CIRS FP3 results and also found smaller \diacety~and \methyla~abundances in 2009 than in 1980 at  $50^{\circ}$N. They suggested that this could be due to variations of the solar activity, as 1980 observations occurred during a solar maximum. Our study shows that \cyano~follows the same trend, which is consistent with this explanation.\\  

                \begin{table*}[!h]
                        \centering
                        \caption{Interannual comparison between abundances of \diacety, \cyano, and \methyla~measured by Voyager I/IRIS in November 1980 \citep{Coustenis1995} and in this study in 2009.}
                        \label{tab_comp_Voyager_Cass}
                        \bgroup
                        \def\arraystretch{1.5}
                        \begin{tabular}{c|cc |cc|cc}
                                Latitude &                      \diacety&                                                                                                                                                                                                 & \cyano&                                                                                                                                                                                          &\methyla & \\
                          &                                     1980 &  2009                                                                                                                                                                         &1980                                         &  2009  & 
                          1980  & 2009\\
                        \hline
                        \hline 
                         $30^{\circ}$S &  $1.3\pm0.3\times 10^{-9}$ & $< 6.1\times 10^{-10}$                                                             &  $< 1.5\times10^{-9}$ & -       &  
                         $4.5\pm0.7\times 10^{-9}$ & $2.3\pm0.2\times10^{-9}$ \\
                         $30^{\circ}$N & $1.2\pm0.4\times 10^{-9}$ &  $2.0\substack{+0.1\\-0.3}\times10^{-9}$         &       $ < 2.0 \times10^{-9}$& $3.0\substack{+0.5\\-0.7}\times10^{-10}$ & 
                         $6.0\pm1.6\times 10^{-9}$&$4.8\substack{+0.3\\-0.4}\times10^{-9}$  \\ 
                         $50^{\circ}$N & $1.5\pm0.2\times 10^{-8}$ & $3.8 \substack{+0.3\\-0.6}\times10^{-9}$  &  $1.5\pm0.2\times 10^{-8}$& $6.5 \substack{+1.0\\-1.5}\times10^{-10}$ &
                         $3\pm0.2\times 10^{-8}$ &$6.1\substack{+0.4\\-0.5}\times10^{-9}$ \\
                         \hline
                        \end{tabular}
                        \egroup
                \end{table*}
        
        \subsection{Influence of abundances at lower pressure levels on the nadir measurements}

                In this paper, we use uniform {a priori} profiles to retrieve \diacety, \cyano, and \methyla. In \cite{Vinatier2015}, the authors measured the vertical profiles of \diacety~and \methyla~using Cassini/CIRS limb observations, and they showed that the vertical gradients of the abundance of these gases can be steep and exhibit strong temporal variations. In fig. \ref{fig_testsand}, we show two examples of vertical profiles of \diacety~and \methyla~measured by \citet{Vinatier2015}. The abundances measured for these two gases at $79^{\circ}$S in September 2014 (blue dashed lines in the left panels) show a strong enrichment at low pressures (enrichment by a factor of 200 between 0.1~mbar and 0.01~mbar), while much weaker vertical variations were measured at $46^{\circ}$N in 2012 (blue dashed lines in the right panels).\\
                
                In order to evaluate how the shape of the vertical profiles of \methyla~and \diacety, and especially how an enrichment in these species at high altitude can affect our results, we retrieve best-fitting scales factors for \methyla~and \diacety, using the limb measurements from \citet{Vinatier2015} as {a priori} profiles. Above the upper sensitivity limit of the limb data, we use a constant vertical profile. Below the lower sensitivity limit of the limb data, we also set the profile to a constant value with pressure until we reach the condensation level, where  the shape of the profile decreases to mimic the condensation of the two considered species.\\

                We perform these retrievals for several nadir datasets acquired at several latitudes and seasons ($72^{\circ}$N in April 2007, $45^{\circ}$N in September 2012, $75^{\circ}$N in April 2014, $70^{\circ}$S in December 2014) using limb profiles measured at close latitudes and times ($70^{\circ}$N in August 2007, $46^{\circ}$N in June 2012, $71^{\circ}$N in January 2015, $79^{\circ}$S in September 2014). In the following paragraph, we discuss the effects of the \diacety~and \methyla\ {a priori} profiles with the largest and the smallest vertical gradients using two observations:                 
                \begin{itemize}
                        \item the retrieval of the nadir data at $45^{\circ}$N observed in September 2012, using the limb profile measured at $46^{\circ}$N in June 2012 as an {a priori} (low vertical gradients); 
                        
                        \item the retrieval of the nadir data at $70^{\circ}$S observed in December 2014, using the limb profile measured at $79^{\circ}$S in September 2014 as an {a priori} (high vertical gradients)   
                \end{itemize}

                \begin{figure}[!h]
                        \centering
                        \includegraphics[width=1\columnwidth]{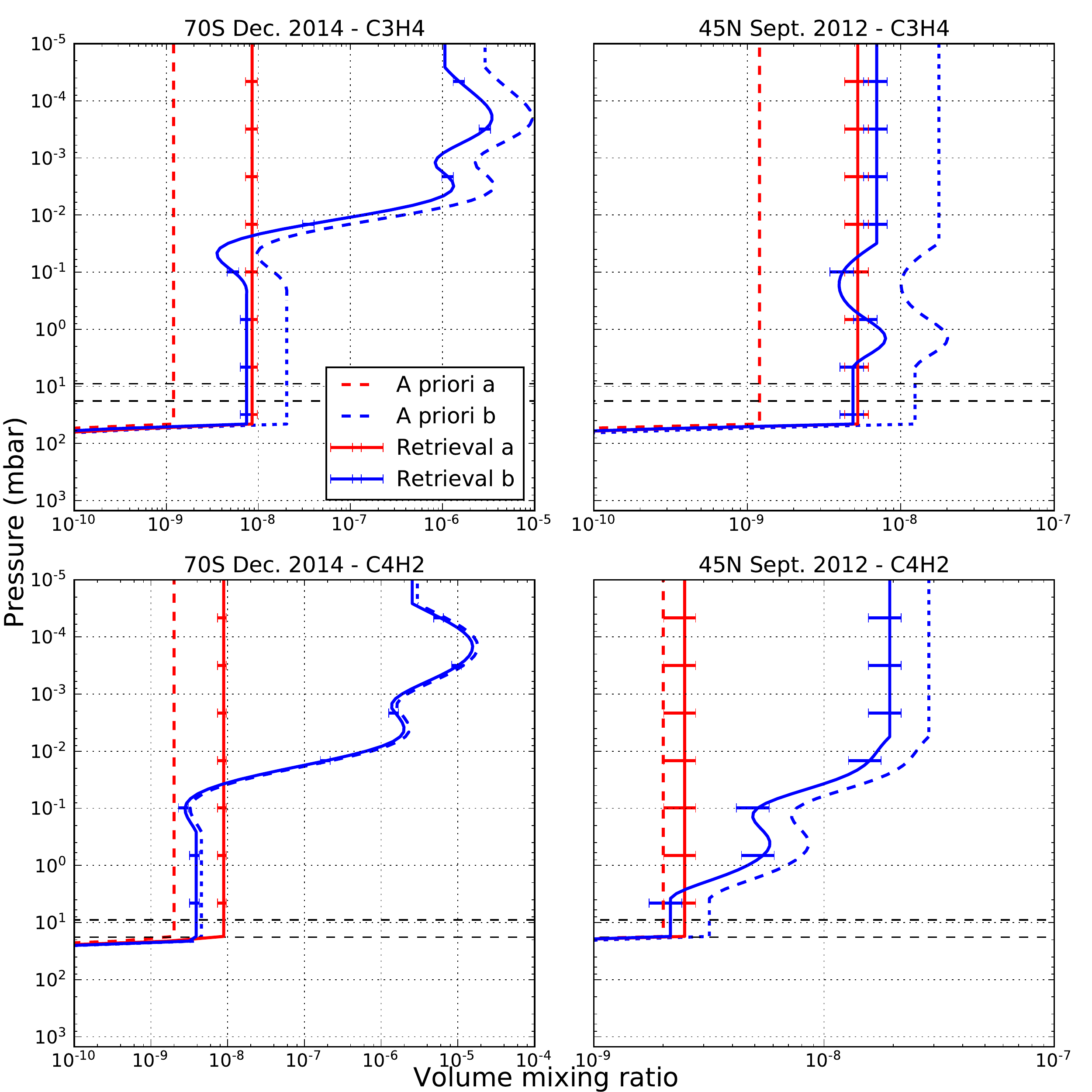}
                        \caption{Vertical profiles of \methyla~(top panels) and \diacety~(bottom panels) retrieved at $70^{\circ}$S in December 2014 (left panels) and $45^{\circ}$N in September 2012 (right panels). Vertical profiles from retrievals \textit{(a)} and their {a priori} are respectively represented by red solid and dashed lines. Vertical profiles from retrievals \textit{(b)} are represented by blue solid lines. {A priori} profiles for retrievals \textit{(b)} are represented by blue dashed lines within the pressure range probed by the CIRS limb observations, and by blue dotted lines outside this pressure range. Thin black dashed lines show the pressure range probed by our CIRS nadir observations. The limb profiles on the left and right panels (blue dashed lines) were respectively measured at $79^{\circ}$S in September 2014 and $46^{\circ}$N in June 2012. At the pressure levels probed by our observations, retrievals \textit{(a)} and \textit{(b)} give consistent results, except for \diacety~at $70^{\circ}$S in December 2014, because of the high vertical gradient of the {a priori} profile \textit{(b)}.}
                        \label{fig_testsand}
                \end{figure}

                \begin{figure}[!h]
                        \begin{center}
                                \includegraphics[width=1\columnwidth]{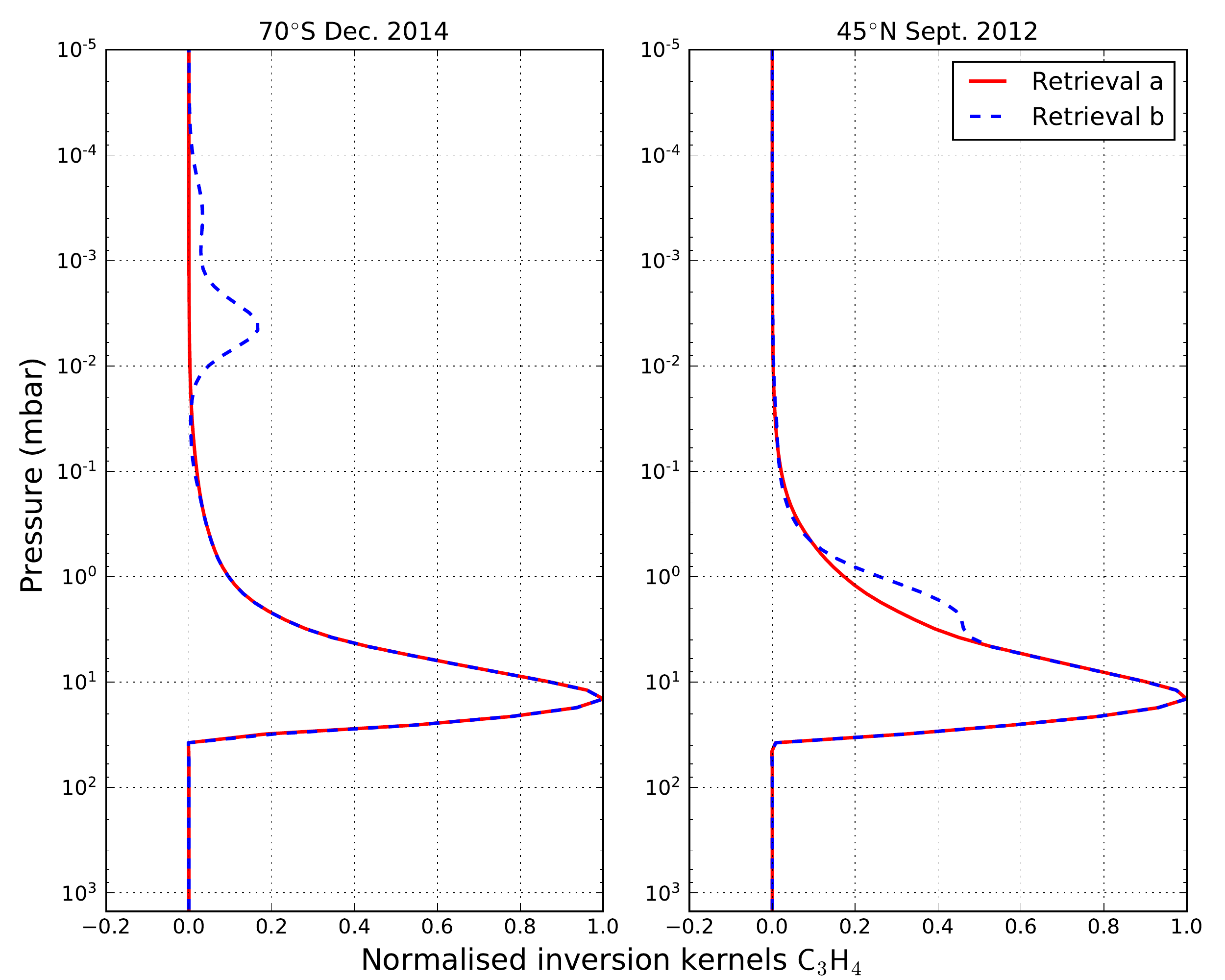}
                                \includegraphics[width=1\columnwidth]{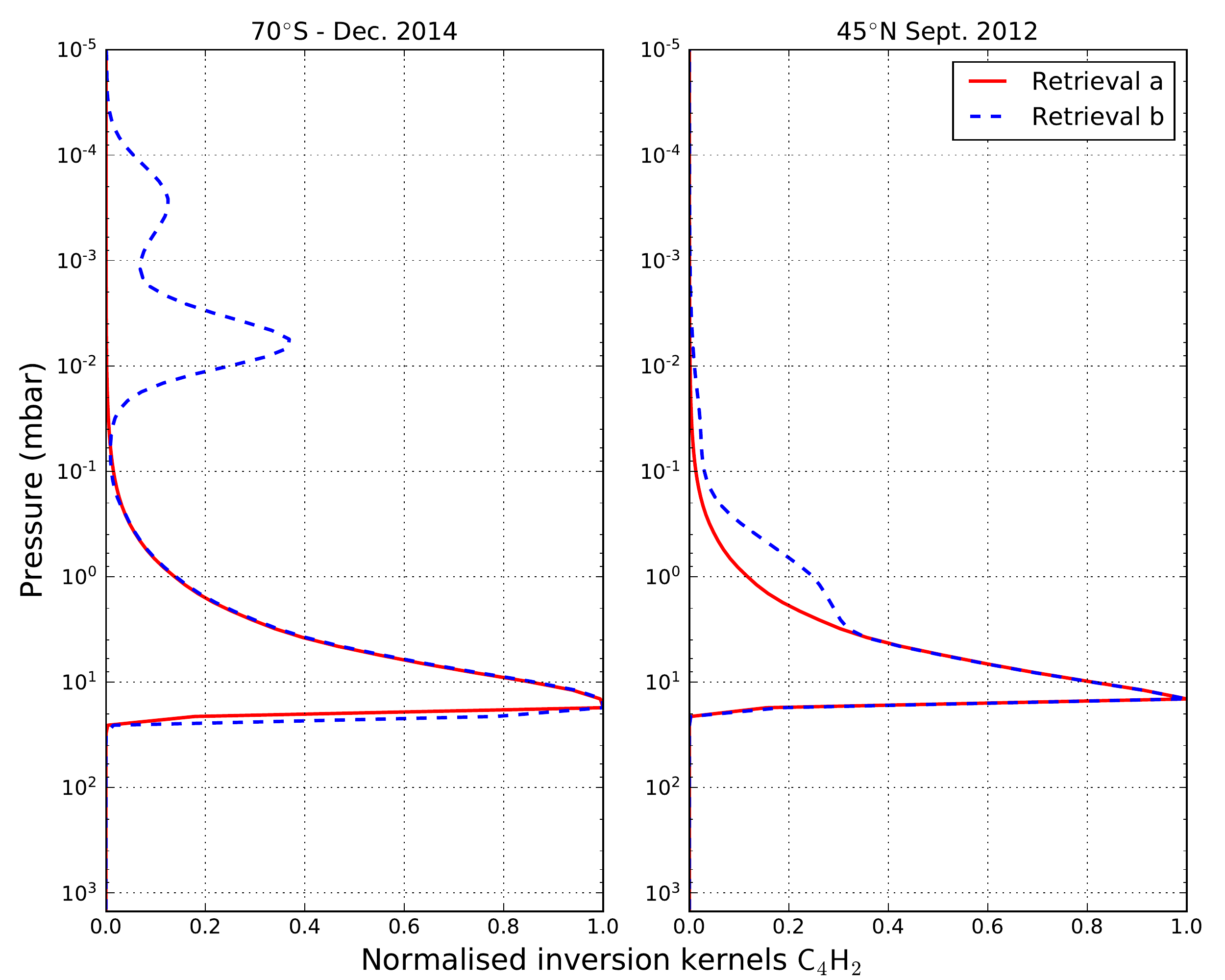}
                                \caption{Inversion kernels for retrievals \textit{(a)} (red solid line) and \textit{b} (blue dashed lines) for \methyla~(top) and \diacety~(bottom). They were calculated from the nadir retrievals performed with the observations at $70^{\circ}$S in December 2014 (left panels) and $45^{\circ}$N in September 2012 (right panels). The vertical gradients of the {a priori} profiles can affect the shape of the contribution functions.}
                                \label{fig_contrib_sand}
                        \end{center}
                \end{figure}

                Figure \ref{fig_testsand} shows the comparison between the results obtained using constant {a priori} profiles (retrievals \textit{(a)}) and {a priori} profiles from the limb observations of \citet{Vinatier2015} (retrievals \textit{(b)}), for \diacety~and \methyla~for the selected examples. Figure \ref{fig_contrib_sand} shows the inversion kernels obtained for \diacety~and \methyla~with the retrievals \textit{(a)} and \textit{(b)}. 
                
                \subsubsection*{Low vertical gradient - $45^{\circ}$N in September 2012}
                
                        When the vertical gradients in the {a priori} profiles of \diacety~and \methyla~are low, the inversion kernels of the retrievals \textit{(a)} and \textit{(b)} are slightly different;  the volume mixing ratios measured at higher altitude are large enough to broaden the base of the contribution functions. However, these variations of the contribution function are small compared to the size of its main lobe. This is consistent with the fact that the profiles obtained from retrievals \textit{(b)} are within error bars from the profiles \textit{(a)} at the pressure levels probed by nadir observations (from 18~mbar to 9~mbar). Above this pressure range, the difference between the profiles \textit{(a)} and \textit{(b)} is larger than the error bars  because we retrieve a scale factor of the {a priori} profiles.
                        
                \subsubsection*{High vertical gradient - $70^{\circ}$S in December 2014}  
                
                        When the vertical gradients of the {a priori} profiles of \methyla~and \diacety~are high, a second lobe can appear in the contribution functions. For \methyla, this second lobe is small compared to the main lobe, which means that the enrichment in \methyla~measured around 0.004~mbar does not affect the the nadir measurements at 15~mbar. For \diacety, the second peak of the contribution function is not negligible compared to the main peak. As a result, the profile \textit{(b)} is significantly lower than the profile \textit{(a)} in the sensitivity range because of the contribution from high altitudes. There is a factor of 2.3  between these two profiles of \diacety, which is the greatest difference found among the different datasets for which we performed these tests. Profiles \textit{(a)} and \textit{(b)} of \methyla~are within error bars.  Thus, we can conclude that the abundances of \diacety~from the nadir measurements at high southern latitudes during autumn may be slightly overestimated, by less than a factor 2.3. However, the relative abundances variations of \diacety~remain robust. \\                
                
                \subsection{Implications for photochemistry and atmospheric dynamics}               
                
                        The seasonal evolution of the abundances of the trace species is the result of a complex interplay between photochemistry and atmospheric dynamics.  Measurements of the trace species distribution at several latitudes and altitudes help to explain these different processes.\\
                
                        Global Climate Models (GCM), \citet{Hourdin1995, Tokano1999, Lebonnois2012}) predict that atmospheric circulation of Titan takes the form of a global cell with an upwelling branch above the summer pole and a subsiding branch above the winter pole. Around equinoxes, as the atmospheric circulation is reversing, there are two Hadley cells with upwelling at the equator and subsidence above the poles. The seasonal evolution of the abundances of \diacety, \cyano, and \methyla~at 15~mbar can be related to the effects of atmospheric dynamics in the lower stratosphere. According to the predictions of the GCMs, high latitudes should be particularly sensitive to the seasonal changes, which is in good agreement with our observations at Titan's south pole. Indeed, in section \ref{sect_res}, we show that after 2012, during southern autumn, high southern latitudes exhibit a strong and sudden enrichment in \cyano, \diacety, and \methyla. This is similar to what has been measured at higher altitudes in the stratosphere by previous studies \citep{Teanby2012, Vinatier2015, Coustenis2016}. This enrichment has been interpreted as the effect of the subsiding branch of the Hadley cell above these latitudes, carrying these photochemical products from the upper stratosphere. Our measurements (see fig. \ref{fig_evmerid}) show that this subsidence above the autumn pole also affects this part of the stratosphere, in good agreement with the atmospheric circulation predicted by the GCM of \citet{Lebonnois2012}. In addition, in \citet{Teanby2012, Vinatier2015}, the authors show that the enrichment in species such as $\mathrm{HCN}$,  \diacety, or \methyla\ appeared in the upper stratosphere at 500~km (0.001~mbar) between June and September 2011, during southern autumn. As the 15 mbar pressure level ($\sim85$~km of altitude) began to exhibit the same enrichment in \diacety, \methyla, and \cyano~between September and November 2012, we can infer that the air enriched in photochemical species has propagated toward the lower stratosphere in one year.\\
                
                        During winter, high northern latitudes were enriched in photochemical species, and contrarily to the southern high latitudes, they exhibit stable abundances of \diacety, \cyano, and \methyla~from northern winter to spring, similarly to what has been observed by \citet{Coustenis2016} for gases such as \diacety, \methyla, or $\mathrm{C_6H_6}$  at 10~mbar. There is a strong difference between the seasonal variations of these species at the probed pressure level (15~mbar) and the seasonal variations measured at lower pressures by \citet{Vinatier2015}. Indeed, they measured a strong enhancement of trace gases between 0.01~mbar and 0.001~mbar, after the equinox (in 2010), which disappeared one year later. Figure \ref{fig_GCM} shows the stream functions predicted by the GCM simulation from \citet{Lebonnois2012} during northern spring. Progressively, the circulation is evolving from two equator-to-pole cells to a single pole-to-pole cell (with upwelling above the north pole and downwelling above the south pole). Figure \ref{fig_GCM} shows that to reach this final state, the northern equator-to-pole cell shrinks in latitude and moves toward the north pole. This small cell may act like a `trap' for the photochemical species, keeping the amount of photochemical products constant near the north pole and preventing them from being advected toward the other latitudes by the bigger cell. The latitudinal extent of the small cell seems to vary with the pressure level: between 5~mbar and 11~mbar it extends from $60^{\circ}$N and $85^{\circ}$N, while it is very narrow at 0.1~mbar as it goes from  $70^{\circ}$N to $75^{\circ}$N. This difference may explain why seasonal variations were measured at low pressures in limb measurements while abundances stay constant in our nadir measurements.\\

                        \begin{figure}[h]
                                \begin{center}
                                        \includegraphics[width=1\columnwidth]{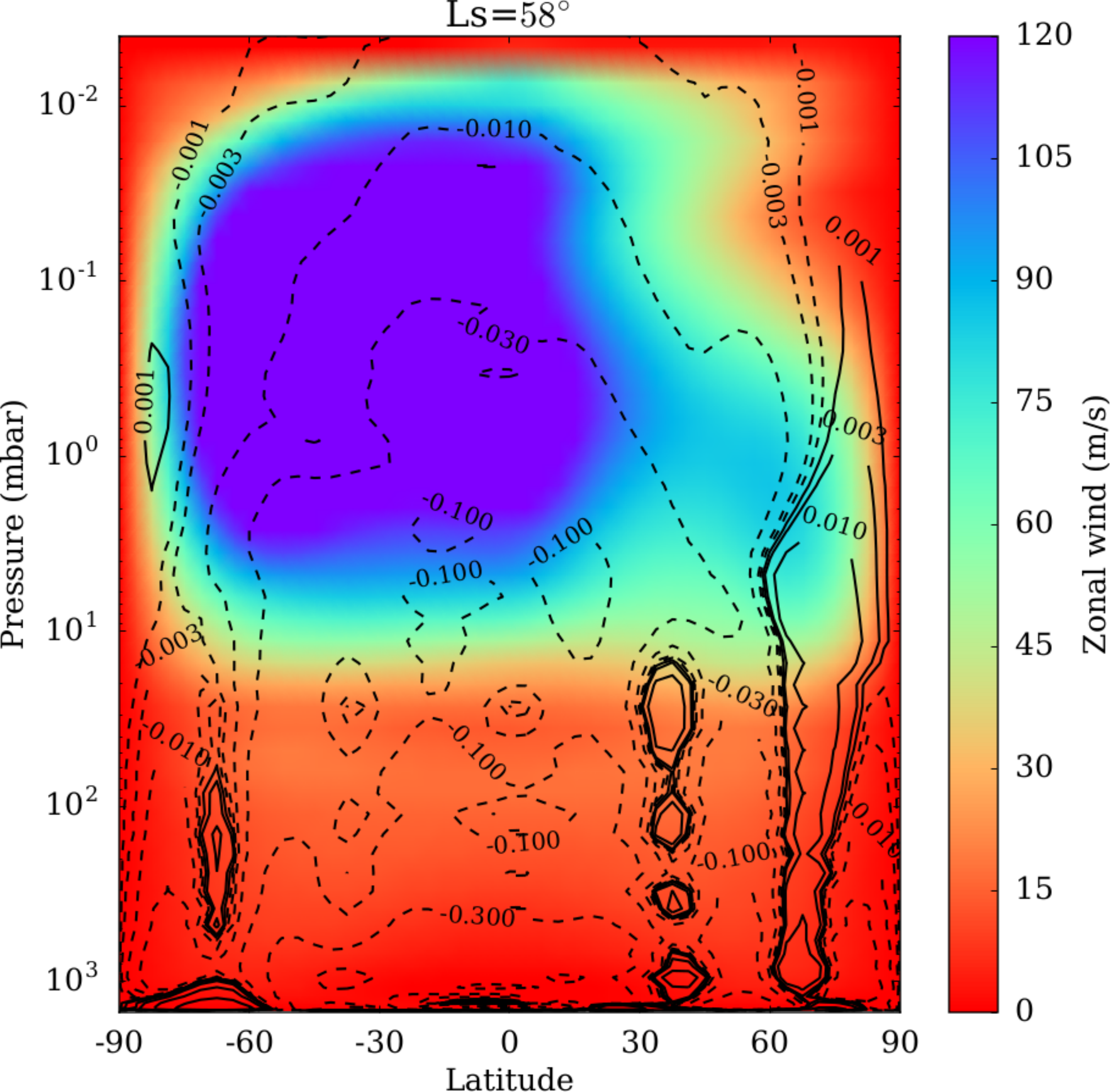}
                                        \caption{Zonal winds and stream function ($10^9$~kg/s) at $L_S=58^{\circ}$ (July 2014) from the numerical simulation in \citet{Lebonnois2012}. Solid and dashed lines indicate respectively clockwise and anticlockwise rotation.}
                                        \label{fig_GCM}
                                \end{center}
                        \end{figure}

                        The seasonal evolution of the poles is also characterised by the strong radiance variations shown in fig. \ref{fig_ev70NS}, centred at $220~\mathrm{cm^{-1}}$. \citet{Jennings2012,Jennings2015} used Cassini/CIRS data to study this spectral feature and suggested (following \citet{Coustenis1999,deKok2008,Anderson2012}) that it is the signature of a cloud mainly composed of nitriles located between 80~km and 150~km. \citet{Jennings2012} suggested that the disappearance of this cloud above the north pole could be related to the increase in insolation at these latitudes which would lead to more photolysis of the source chemicals of this cloud at higher altitudes, and to a lessening of condensation, or to a weakening of the subsiding branch above the north pole. However, during spring, abundances of various gases have stayed constant (in this study and in \citet{Coustenis2016}) in the lower stratosphere above high northern latitudes. This would imply that the disappearance of the cloud is more related to the seasonal insolation variations than to dynamics. Above the south pole, we show that the emission at $220~\mathrm{cm^{-1}}$ has kept on increasing in 2014 and 2016, which is quite surprising as \citet{Jennings2015} suggested that a decrease in the radiances at these latitudes should occur in 2015-2016, based on north pole behaviour.\\

                        In  this study, we find that meridional distributions of \diacety~and \methyla~do not evolve between $50^{\circ}$S and $30^{\circ}$N from 2006 to 2015. This is compatible with the results of \citet{Bampasidis2012}, where the distributions of these gases were measured from 2006 to 2012. \citet{Bampasidis2012} also monitored sharp variations in \methyla~and \diacety~(and other gases such as $\mathrm{C_2H_4}$ or $\mathrm{HCN}$) between 2008 and 2010, characterised by a steep increase until mid-2009 and then a decrease in their abundances. They attributed this evolution to the rapid changes in atmospheric circulation around equinox, and especially the weakening of the vortex observed in the northern winter hemisphere. These temporal variations are not present in our data.  It may be because we do not have enough data at $50^{\circ}$N in 2009-2010. It might also indicate that unlike  higher pressure levels,  the atmospheric circulation at 15 mbar does not exhibit sharp changes around the equinox, and evolves more steadily.\\

                        Photochemistry also controls the distribution of trace gases in the stratosphere. In fig. \ref{fig_evmerid}, we show that at mid-northern latitudes \methyla~abundances are constant, while \cyano~and \diacety~abundances decrease in 2014-2016. The photochemical model of \citet{Lavvas2008a} predicts that \diacety~photochemical lifetime at 150~km (2~mbar) is 0.2~yr, which is 10 times lower than for \methyla~(2~yr). If there is a similar difference between the photochemical lifetimes of these two species at 15~mbar, this can explain why a diminution of \diacety~abundances is observed, while \methyla~abundances do not vary. As \cyano~follows the same trend as \diacety, this would suggest that \cyano~and \diacety~photochemical lifetimes at 15~mbar are of the same order of magnitude.\\
                
                        Photochemical models disagree on the loss mechanisms of these species. In the photochemical model of Titan's atmosphere presented by \citet{Krasnopolsky2014}, photolysis is the major sink for \cyano~(68\% of loss), whereas it is a minor loss mechanism for \diacety~and \methyla~(7\% and 9\% of loss respectively). This would mean that  \cyano~could be more sensitive than \diacety~to the seasonal variations of insolation, which is not consistent with our results. However, in the photochemical model of \citet{Vuitton2014}, photolysis is a minor loss reaction for the three studied species at the probed pressure-level. Their main loss reaction is the combination with atomic hydrogen, for instance, 
                                \begin{equation*}
                                        \mathrm{C_4H_2} + \mathrm{H} \rightarrow \mathrm{C_4H_3} + h\nu,
                                \end{equation*}
                \noindent which would be consistent with the fact that \cyano~and \diacety\ vary in a similar way in the northern hemisphere during spring.\\  
                
\section{Conclusion}
                
        In this work, we study the seasonal evolution of Titan's lower stratosphere, using Cassini/CIRS far-infrared observations. These data allow us to probe the atmosphere around the 15~mbar pressure level and to measure the abundances of three photochemical by-products: \cyano, \methyla, and \diacety. Thanks to the long duration of the Cassini mission and the good latitudinal coverage of these data, we have been able to monitor the evolution of these species over the whole latitude range from 2006 to 2016, i.e. from northern winter to mid-spring.\\
                
        The most striking feature is the asymmetry in the seasonal evolution of high northern latitudes where the volume mixing ratios of \diacety, \cyano, and \methyla~have stayed approximately constant from northern winter to spring, whereas high southern latitudes  exhibit a sudden and strong enrichment in these species during southern autumn, consistent with the observations at 10~mbar of \citet{Coustenis2016}. We also show that \methyla~has a  different seasonal evolution compared to \cyano~and \diacety~at mid-northern latitudes, which may be due to shorter photochemical lifetimes for the two latter species.\\
                
        The evolution of the high latitudes is consistent with the seasonal evolution of Titan's atmospheric circulation predicted by the GCM of \citet{Lebonnois2012} as the effect of a subsidence above the south pole and the presence of a small circulation cell towards the high northern latitudes can explain our results.\\

\section*{Acknowledgements}

        The authors thank Véronique Vuitton for very useful discussions about Titan's photochemistry and Emmanuel Lellouch for his comments about possible effects of high altitudes abundances on our retrievals. We also thank the anonymous reviewer for the suggestions that improved this paper. This research was funded by the UK Sciences and Technology Facilities Research council (grant number ST/MOO7715/1) and the Cassini project.\\
                        
\bibliographystyle{aa} 
\bibliography{biblio.bib} 

\setcounter{table}{0}

\longtab[!H]{
                                        \begin{longtable}{lllllll}
                                                
                                                \caption{\label{table_obs}Datasets used in this study. N FP1 and N FP4 respectively stand for the number of spectra measured with FP1 and  FP4 during the acquisition. FOV is the field of view. The asterisk denotes the datasets which have already been presented in \citet{Teanby2009}. The symbol \dag~denotes datasets for which we perform the retrievals using small domains around the spectral bands of \diacety, \cyano, and \methyla, as described in section \ref{sect_conti}.}\\
                                                \hline
                                                \hline
                                                Observations  & Date  & N FP1 & Lat. FP1($^{\circ}$N) & FOV FP1($^{\circ}$)& N FP4 & Lat. FP4($^{\circ}$N)\\
                                                \hline
                                                \endfirsthead
                                                \caption{continued.}\\
                                                \hline\hline
                                                Observations  & Date  & N FP1 & Lat. FP1($^{\circ}$N) & FOV FP1($^{\circ}$)& N FP4 & Lat. FP4($^{\circ}$N)\\
                                                \hline
                                                \endhead
                                                \hline
                                                \endfoot
                                                CIRS\_036TI\_FIRNADCMP002\_PRIME* & 28 Dec. 2006 &  136 & -89.1 & 12.6 & 684 & -74.2 \\ 
                                                CIRS\_036TI\_FIRNADCMP003\_PRIME*\dag & 27 Dec. 2006 &  321 & 78.6 & 21.0 & 1620 & 74.2 \\ 
                                                CIRS\_037TI\_FIRNADCMP001\_PRIME\dag & 12 Jan. 2007 &  161 & 75.2 & 19.1 & 815 & 83.3  \\
                                                CIRS\_037TI\_FIRNADCMP002\_PRIME & 13 Jan. 2007 &  107 & -70.3 & 20.6 & 540 & -84.2 \\
                                                CIRS\_038TI\_FIRNADCMP001\_PRIME\dag & 28 Jan. 2007 & 254 & 86.3 & 16.7 & 1275 & 69.0 \\
                                                CIRS\_040TI\_FIRNADCMP001\_PRIME* & 09 Mar. 2007 &  159 & -49.2 & 21.1 & 795 & -61.2\\
                                                CIRS\_040TI\_FIRNADCMP002\_PRIME*\dag& 10 Mar. 2007 &  109 & 88.8 & 13.3 & 545 & 69.0 \\
                                                CIRS\_041TI\_FIRNADCMP002\_PRIME* & 26 Mar. 2007 &  102 & 61.2 & 19.3 & 525 & 47.9 \\
                                                CIRS\_042TI\_FIRNADCMP001\_PRIME* & 10 Apr. 2007 &  103 & -60.8 & 26.0 & 515 & -78.0 \\
                                                CIRS\_043TI\_FIRNADCMP002\_PRIME*\dag & 27 Apr. 2007 &  104 & 77.1 & 20.0 & 535 & 55.0  \\
                                                CIRS\_044TI\_FIRNADCMP002\_PRIME* & 13 May 2007 &  104 & -0.5 & 18.8 & 524 & -17.3  \\
                                                CIRS\_045TI\_FIRNADCMP002\_PRIME* & 29 May 2007 &  346 & 52.4 & 29.5 & 1735 & 30.0  \\
                                                CIRS\_046TI\_FIRNADCMP002\_PRIME & 14 Jun. 2007 &  102 & -20.8 & 19.0 & 510 & -38.3 \\
                                                CIRS\_052TI\_FIRNADCMP002\_PRIME* & 19 Nov. 2007 &  272 & 40.3 & 26.5 & 1365 & 19.3  \\
                                                CIRS\_053TI\_FIRNADCMP001\_PRIME* & 04 Dec. 2007 &  223 & -40.2 & 25.8 & 1119 & -49.8 \\
                                                CIRS\_055TI\_FIRNADCMP001\_PRIME & 05 Jan. 2008 &  190 & 18.7 & 30.5 & 960 & 5.6  \\
                                                CIRS\_055TI\_FIRNADCMP002\_PRIME & 06 Jan. 2008 &  284 & 44.6 & 22.2 & 1420 & 39.9  \\
                                                CIRS\_069TI\_FIRNADCMP002\_PRIME* & 28 May 2008 &  112 & 9.5 & 19.3 & 565 & -9.2 \\
                                                CIRS\_107TI\_FIRNADCMP002\_PRIME & 27 Mar. 2009 &  164 & 33.5 & 30.4 & 821 & 50.1  \\
                                                CIRS\_110TI\_FIRNADCMP001\_PRIME & 06 May 2009 &  282 & -68.1 & 25.7 & 1410 & -59.4  \\
                                                CIRS\_111TI\_FIRNADCMP002\_PRIME & 22 May 2009 &  168 & -27.1 & 23.1 & 842 & -21.6 \\
                                                CIRS\_112TI\_FIRNADCMP001\_PRIME & 06 Jun. 2009 &  218 & 48.7 & 21.0 & 1090 & 59.0  \\
                                                CIRS\_112TI\_FIRNADCMP002\_PRIME & 07 Jun. 2009 &  274 & -58.9 & 20.2 & 1370 & -39.9  \\
                                                CIRS\_160TI\_FIRNADCMP002\_PRIME & 30 Jan. 2012 &  280 & -0.2 & 18.3 & 1474 & 6.9  \\
                                                CIRS\_172TI\_FIRNADCMP001\_PRIME & 26 Sep. 2012 &  282 & 44.9 & 18.5 & 1410 & 50.9  \\
                                                CIRS\_172TI\_FIRNADCMP002\_PRIME\dag & 26 Sep. 2012 &  270 & -70.4 & 23.2 & 1352 & -70.8  \\
                                                CIRS\_174TI\_FIRNADCMP002\_PRIME & 13 Nov. 2012 &  298 & -71.8 & 21.8 & 1493 & -52.9  \\
                                                CIRS\_194TI\_FIRNADCMP001\_PRIME & 10 Jul. 2013 &  186 & 30.0 & 19.7 & 935 & 46.1  \\
                                                CIRS\_197TI\_FIRNADCMP001\_PRIME & 11 Sep. 2013 &  330 & 60.5 & 19.4 & 1650 & 64.0 \\
                                                CIRS\_198TI\_FIRNADCMP001\_PRIME & 13 Oct. 2013 &  187 & 88.9 & 8.7 & 935 & 72.9  \\
                                                CIRS\_198TI\_FIRNADCMP002\_PRIME\dag & 14 Oct. 2013 &  306 & -69.8 & 24.0 & 1530 & -85.8 \\
                                                CIRS\_199TI\_FIRNADCMP001\_PRIME & 30 Nov. 2013 &  329 & 68.4 & 23.9 & 1650 & 85.5  \\
                                                CIRS\_200TI\_FIRNADCMP001\_PRIME & 01 Jan. 2014 &  187 & 49.9 & 19.6 & 935 & 37.2 \\
                                                CIRS\_201TI\_FIRNADCMP001\_PRIME & 02 Feb. 2014 &  329 & 19.9 & 26.8 & 1649 & 27.6 \\
                                                CIRS\_203TI\_FIRNADCMP001\_PRIME & 07 Apr. 2014 &  187 & 75.0 & 18.0 & 935 & 67.7 \\
                                                CIRS\_207TI\_FIRNADCMP001\_PRIME\dag & 20 Aug. 2014 &  179 & -70.0 & 17.8 & 895 & -79.4  \\
                                                CIRS\_207TI\_FIRNADCMP002\_PRIME & 21 Aug. 2014 &  39 & 79.8 & 16.1 & 195 & 78.1  \\
                                                CIRS\_208TI\_FIRNADCMP002\_PRIME & 22 Sep. 2014 &  175 & 60.5 & 17.8 & 875 & 75.6  \\
                                                CIRS\_210TI\_FIRNADCMP001\_PRIME\dag & 10 Dec. 2014 & 329 & -70.3 & 25.5 & 1646 & -50.5  \\
                                                CIRS\_218TI\_FIRNADCMP001\_PRIME & 06 Jul.  2015 &  249 & -20.0 & 19.9 & 1250 & -2.9 \\
                                                CIRS\_222TI\_FIRNADCMP002\_PRIME & 29 Sep. 2015 &  233 & -0.1 & 18.6 & 530 & 18.3 \\
                                                CIRS\_232TI\_FIRNADCMP001\_PRIME & 16 Feb. 2016 &  249 & -50.2 & 24.5 & 1245 & -31.0 \\
                                                CIRS\_235TI\_FIRNADCMP001\_PRIME\dag & 06 May 2016 &  163 & -60.0 & 19.7 & 820 & -45.1 \\
                                                CIRS\_236TI\_FIRNADCMP001\_PRIME \dag & 07 Jun. 2016 &  88 & -70.0 & 15.8 & 440 & -53.2 \\
                                                CIRS\_236TI\_FIRNADCMP002\_PRIME & 07 Jun. 2016 &  238 & 60.0 & 19.8 & 1193 & 78.7  \\
                                \end{longtable}
                              }

\end{document}